\documentclass[fleqn,usenatbib]{mnras}
\usepackage{fix-cm}
\usepackage{blindtext}

\usepackage[T1]{fontenc} 
\usepackage{newtxtext,newtxmath}
\usepackage[dvipsnames]{xcolor}
\usepackage{gensymb}
\usepackage{multirow}
\usepackage{bm}
\usepackage{verbatim}
\usepackage{amsmath}
\usepackage{siunitx}
\usepackage{graphicx}
\usepackage{hyperref}
\usepackage{cleveref}
\newcommand{\be}{\begin{equation}}
\newcommand{\ee}{\end{equation}}
\newcommand{\nn}{\nonumber}
\newcommand{\bea}{\begin{eqnarray}}
\newcommand{\eea}{\end{eqnarray}}
\renewcommand{\d}{{\rm d}}

\newcommand{\hi}{\textrm{H\textsc{i}}}

\DeclareMathOperator\erf{erf}
\DeclareRobustCommand{\VAN}[3]{#2}
\let\VANthebibliography\thebibliography
\def\thebibliography{\DeclareRobustCommand{\VAN}[3]{##3}\VANthebibliography}

\everymath{\displaystyle}

\title[Radio-GW synergies]{Exploring future synergies for large-scale structure between gravitational waves and radio sources}
\author[S. Zazzera et al.]{
Stefano Zazzera,$^{1}$\thanks{E-mail: s.zazzera@qmul.ac.uk}
José Fonseca,$^{2,3,4}$
\thanks{E-mail: jose.fonseca@astro.up.pt}
Tessa Baker$^{5}$
and Chris Clarkson$^{1,3}$
\\
$^{1}$Department of Physics \& Astronomy, Queen Mary University of London, Mile End Road, London E1 4NS, United Kingdom\\
$^{2}$Instituto de Astrofisica e Ci\^{e}ncias do Espa\c{c}o, Universidade do Porto CAUP, Rua das Estrelas, PT4150-762 Porto, Portugal\\
$^{3}$Department of Physics \& Astronomy, University of the Western Cape, Cape Town 7535, South Africa\\
$^{4}$Departamento de F\'isica e Astronomia, Faculdade de Ci\^{e}ncias, Universidade do Porto, Rua do Campo Alegre 687, PT4169-007 Porto, Portugal\\
$^{5}$Institute of Cosmology and Gravitation, University of Portsmouth, Burnaby Road, Portsmouth PO1 3FX, UK
}
\date{Accepted XXX. Received YYY; in original form ZZZ}

\pubyear{\the\year{}}

\begin{document}
\label{firstpage}
\pagerange{\pageref{firstpage}--\pageref{lastpage}}
\maketitle

\begin{abstract}
Future third-generation gravitational wave detectors like the Einstein Telescope (ET) and Cosmic Explorer (CE) are expected to detect millions of binary black hole (BBH) mergers.   
Alongside these advances, upcoming radio surveys, such as the Square Kilometer Array Observatory (SKAO) will provide new sets of cosmological tracers. These include mapping the large-scale distribution of neutral hydrogen (\hi) using intensity mapping (IM) and \hi\ and radio continuum galaxies. In this work, we will investigate synergies between gravitational waves (GW) and radio tracers through a multi-tracer approach. We first forecast the precision on the clustering bias of GWs by cross-correlating data from an ET-like detector with an SKAO IM survey. Our results indicate that this approach can constrain the GW clustering bias to within $2\%$ up to $z = 2.5$. Additionally, we explore the potential of a triple cross-correlation using GWs, IM, and photometric galaxies from a survey like the Vera Rubin Observatory's Legacy Survey of Space and Time (LSST). This multi-tracer method enhances constraints on the magnification lensing effect, achieving percent-level precision, and allows for a measurement of the Doppler effect with approximately $15\%$ uncertainty. Furthermore we show for the first time that this method could achieve the precision required to measure subdominant gravitational potential contributions to the relativistic corrections, which had thought to be below cosmic variance. Our analysis highlights the potential of cross-correlations between GWs and radio tracers to improve constraints on astrophysical properties of BBHs, measure relativistic effects, and perform null tests of GR in cosmological scales.
\end{abstract}


\section{Introduction}\label{sec:intro}

Gravitational waves (GWs) have recently gained traction in the literature as potentially powerful cosmological probes \citep{Mastrogiovanni_24, Mastrogiovanni_2020, 2017NatCo...8.1148L,Auclair_2023,Cai_2017,2018FrASS...5...44E}.
Since the first detection by the LIGO-Virgo-KAGRA collaboration (LVK) in 2015 \citep{GWfirst}, numerous studies have demonstrated their potential to constrain cosmological parameters and test General Relativity \citep{2025PhRvD.112h4080A, Palmese:2025zku, Abbott_2023, Chen:2023wpj, Mastrogiovanni_24, Ezquiaga:2021ayr, Mancarella:2021ecn, Leyde:2022orh, Palmese:2021mjm, Finke:2021aom, DES:2020nay}. The growing number of detections, with observing runs like O4 and O5, and the future advent of third-generation detectors such as the Einstein Telescope (ET) \citep{Punturo_2010} and Cosmic Explorer (CE) \citep{Evans_2021}, will provide a very large catalog of binary black hole (BBH) merger events. ET alone is projected to detect around $10^6$ BBHs over a ten-year period, a large increase in available data \citep{BlueBook, Iacovelli:2022bbs, Branchesi:2023mws}.

In parallel, the next decade will witness significant advancements in radio astronomy, particularly with the Square Kilometer Array Observatory (SKAO) \citep{2020PASA...37....7S}. From the several surveys to be conducted by the SKAO, one will be able to trace the Large-scale structure of the Universe with three distinctive tracers of dark matter: \hi\ intensity mapping (IM), \hi\ spectroscopic galaxy surveys and radio continuum galaxy surveys. \hi\ corresponds to the line emission due to the spin-flip transition between the two hyperfine states of neutral hydrogen with a wavelength of $21$ cm. One can exploit the redshifted \hi\ line to perform spectroscopic galaxy surveys \citep{Santos:2015G/}. In the absence of other cosmological emission lines in the radio frequencies one has a one-to-one relationship between observed frequency and redshift of \hi\ emission. One can then use Intensity Mapping (IM) \citep{2017MNRAS.464.1948F}, whereby the \hi\ brightness temperature is measured in \emph{voxels} of the universe without detecting the individual galaxies. This way fluctuations in the brightness temperature trace fluctuations in the dark matter distribution. 
In addition, some radio galaxies do not present any emission line, only emission in the continuum. While getting redshift information is limited one can still perform cosmological tests with such catalogues \citep{2021MNRAS.506.4121A}. The Medium-Deep Band 2 Survey will cover $5,000 \,\text{deg}^2$ with approximately $10,000$ hours of observation, providing an \hi\ galaxy redshift survey out to $z \sim 0.4$. Meanwhile, the Wide Band 1 Survey will span $20,000 , \text{deg}^2$ over the same observation time, performing \hi\ intensity mapping from $z \sim 0.35$ to $z \sim 3$. Future extensions like SKAO2 will extend \hi\ galaxy surveys to $z \sim 2$, creating further opportunities for cosmological analyses \citep{Yahya_2015}.

Cross-correlating GWs with radio tracers of large-scale structure offers a promising avenue for investigating both cosmology and the properties of BBH populations. By comparing the clustering of GWs with \hi\ galaxies, \hi\ IM and continuum galaxies, we can measure the GW clustering bias and explore the large-scale distribution of matter. In particular, a previous study examined such cross-correlations to study the origin of the BBHs which produced the GWs by focussing on the clustering bias \citep{Scelfo_2022_2}. They showed that this kind of analysis will be able to distinguish between black hole mergers of astrophysical or primordial origin. Further, \citet{Pedrotti_2025} carried out an extensive forecast analysis on the measurement of the GW clustering bias when cross-correlated with Euclid data, work also explored by \citet{Dehghani_2025}, although with a different galaxy catalog.

This paper builds upon previous work that examined cross-correlation with GWs \citep{Afroz:2024joi,2021PhRvD.103d3520M,Mukherjee:2019wcg,Mukherjee:2019wfw,Mukherjee:2022afz,mukherjee2018}, for example with galaxy surveys \citep{Libanore_2021, Scelfo_2020, 2024JCAP...02..023B, Pedrotti_2025,2025arXiv251008699S,2025arXiv251019931P} and intensity mapping \citep{Scelfo_2022_2}. In particular, we extend the analysis we presented in \citet{2025MNRAS.537.1912Z} where we produced forecasts on different combinations of GWs detectors, and current and future galaxy surveys such as LSST \citep{LSST_2012,LSST_2021,LSST_book} by incorporating SKAO radio surveys as well. Here we are following a framework set up in \citet{2023JCAP...08..050F,Namikawa,2024JCAP...02..023B} where the clustering of GWs is treated in luminosity distance space  (LDS). In fact, unlike galaxy surveys or intensity mapping, GWs from BBH mergers do not carry information on the redshift of the source, but rather a $D_L$ measurement; that is, unless an electromagnetic counterpart is detected. It is worth noting that one can also use the correlations between GWs and galaxy surveys as a standard ruler, as shown by \citet{2025JCAP...04..008F}. The sole height of the cross-correlation becomes such a ruler which we can then use to build an Hubble diagram. This in turn is used to constrain the amount of matter and the Hubble constant.

Furthermore, the clustering of GWs in LDS is affected by three important parameters, similarly to what happens in redshift space (RS) for other tracers like galaxies. The GW clustering bias describes how GWs trace the underlying dark matter distribution, analogous to the bias of galaxy tracers. 
Additionally, the limited sensitivity of GW detectors introduces observational effects captured by the magnification bias \citep{Maartens_2021}. The magnification bias $s^{GW}$ gauges the amplitude of corrections to the number density contrast arising from gravitational lensing, which can enhance or diminish the observed number of GWs by magnifying sources into or out of detectability.
The evolution bias $b_e^{GW}$, on the other hand, accounts for the redshift evolution of the intrinsic merger rate and the detector's ability to observe the cosmic evolution of the sources. Further analysis and modelling of the magnification and evolution biases of GWs is described in \citet{Zazzera_2024}. These parameters are directly linked to the properties of the BBH population producing the GWs observed, and so measuring them would imply constraining these properties. In particular, $s^{GW}$ and $b_e^{GW}$ depend on the intrinsic merger rate and chirp mass distribution, whilst measuring the clustering bias would directly link the distribution of BBHs to that of the underlying dark matter in the Universe, potentially acting as probe to differentiate between black holes living in dark matter halos and those in isolation, whether of primordial origin or else \citep{Scelfo_2018,Scelfo_2020}.

First we look at how combining data from different GW detector and radio surveys can improve constraints on the GW clustering bias. Then, focussing on cross-correlating ET-like data with an IM survey, we produce forecasts on the detection of different relativistic effects that affect the observed number count fluctuation and thus the angular power spectrum, namely the lensing correction, luminosity distance space distortions (LSDs) and Doppler term. Finally, we introduce a novel approach using a three-tracer correlation between GWs from an ET-like detector, \hi\ intensity mapping from an SKA1-MID-like survey, and galaxies from an LSST-like survey to forecast constraints on clustering, magnification and evolution biases of GWs and on the measurements of the relativistic effects.

The paper is structured as follows. In \autoref{sec:setup}, we outline the theoretical framework describing the large-scale clustering of GWs and radio tracers. Then, \autoref{sec:surveys} provides an overview of the observational characteristics of SKAO's tracers and the GW datasets used in this analysis, while \autoref{sec:fisher} presents the Fisher formalism adopted to forecast parameter constraints. In \autoref{sec:clusbias}, we present our results for the clustering bias of GWs and in \autoref{sec:relativistic} we explore the detection of relativistic effects. Further, \autoref{sec:triple} introduces the triple-correlation analysis using ET, \hi\ IM, and LSST. Finally, we summarize our conclusions in \autoref{sec:conclusions}.

\section{Number Count Fluctuation}\label{sec:setup}
In this section we briefly summarise the relevant equations describing the number counts (or density contrast) and the relativistic effects in both luminosity distance and redshift space. However, a more complete description is either found in Appendix \ref{sec:amplitudes} or in \citet{2025MNRAS.537.1912Z}.

We firstly define the linearly perturbed line element in the conformal Newtonian gauge, used to derive the results listed below:
\be\label{eq:metric}
ds^2=a^2(\eta)\left[-(1+2\Psi)d\eta^2+(1-2\Phi)\delta_{ab}dx^a dx^b\right]\,,
\ee
where $\Phi$ and $\Psi$ are the metric potentials and $\eta$ is conformal time.

The observed density contrast for tracer A, measured through the distance type $X$ (i.e. either redshift $z$ or luminosity distance $D_L$) can be expressed in terms of different contributions as \citep{Bonvin_2008,Challinor_2011,2023JCAP...08..050F}:
\bea\label{eq:coefficients}
\Delta(\boldsymbol{n},X) &=& b\delta_n+A^X_{GRV}\partial_r(\bm v\cdot\bm n) +\int_0^{\bar r} \d r\ \frac{A^X_{L}}{\bar r}\Delta_{\Omega}(\Phi+\Psi) \nn\\
&& +A_D^X(\bm v\cdot\bm n)+ g^{X}(\Psi,\Phi,r) \, . \
\eea
In order, we have the matter density contrast with $\delta_n$ being the density contrast in Newtonian gauge and $b$ the clustering bias of the tracer in question; then follows distortions from the gradient of radial velocity, lensing magnification, Doppler term and the final term, $g^X(\Psi,\Phi,r)$ is a function grouping together further subdominant relativistic corrections (consisting of the metric potentials and related derivatives and integrals), with $r$ being the comoving position $r = \int{\rm{d}}z'/H$. Different tracers are measured through a different distance measure; for instance galaxies are redshift tracers, whereas GWs are measured in luminosity distance space. Thus, depending on which tracer is used, the amplitudes of each contribution to the number counts have different forms. An interested reader can find the full expression for each in both redshift space (RS) and luminosity distance space (LDS) in Appendix \ref{sec:amplitudes}.

Notably, $A_L$ and $A_D$ (and several terms inside $g^X(\Psi,\Phi,r)$) are found to be dependent on two other tracer-specific parameters, $s$ and $b_e$. These are, respectively, the magnification and evolution biases of a given tracer. An in-depth analysis of these parameters for galaxies and intensity mapping is presented in \citet{Maartens_2021}, and for GWs in \citet{Zazzera_2024}. 

From here, one can expand these fields into spherical harmonics and construct the angular power spectrum.
As the data is usually binned into distance intervals, we express the angular power spectrum for the $i$th and $j$th bin as:
\be\label{eq:cl_gen}
C^{ij}_\ell= 4\pi \int {\rm d}\ln k\ \Delta^{A,i}_{\ell}(k)\Delta^{B,j}_{\ell}(k)\ {\cal P}_{\cal R}(k)\,,
\ee
where ${\cal P}_{\cal R}(k)$ is the power spectrum of primordial perturbations and $\cal R$ is the curvature perturbation. This is effectively cross-correlating counts of tracer $A$ in the $i$th bin with those of tracer $B$ in bin $j$. The functions $\Delta^{A,i}_{\ell}(k)$ include the transfer functions needed to relate the primordial perturbations to the contributions to the observed number counts in \autoref{eq:coefficients}, and can be found in \citet{2023JCAP...08..050F} for a tracer in luminosity distance space.


\section{Surveys}\label{sec:surveys}
In this section we provide the properties of the surveys and tracers used, in both the radio and GW domains. We address expressions for the number densities, and clustering, magnification and evolution biases.

\subsection{Radio tracers}
We consider three different radio tracers in this paper: \hi\ galaxies, radio continuum galaxies and \hi\ intensity mapping. 

For the former two the approach is almost identical to the one taken in \citet{2025MNRAS.537.1912Z}. 
We start by modelling the number density of observed sources for \hi\ galaxies, which is dependent on the flux density threshold $S_{\rm rms}$ and on the observed line profile, i.e. how well the hydrogen spectral line is resolved. The resulting flux threshold becomes \citep{Maartens_2021}: 
\bea\label{eq:srms}
S_c(z) = S_{\rm rms}(z)\frac{N_{ \rm cut}}{10} \, .\
\eea
In order to model the number density of \hi\ galaxies we follow the results from \citet{Yahya_2015}, where the authors used the S$^3$-SAX simulation \citep{Obreschkow_2009}. Here, each galaxy has a redshift, \hi\ luminosity and line profile, allowing for the extraction of the following fitting formula as a function of redshift and detection threshold:
\bea\label{eq:ngal_hi}
N_g^{\hi}(z,S_c) &=& 10^{c_1(S_c)}z^{c_2(S_C)}\exp{[-c_3(S_c)z]} \quad \rm{deg}^{-2} \, .
\eea
The parameters $c_1,c_2$ and $c_3$ depends on the flux sensitivity $S_c$ used, and are reported in Table 3 of \citet{Yahya_2015}. We adopt $S_c=100$ for SKAO1, and $S_c=5$ for SKAO2 \citep{Yahya_2015,Santos:2015G/}. 

The clustering bias of \hi\ galaxies is calculated with the same simulation, resulting in:
\bea\label{eq:b_HIgal}
b^{\hi, \rm gal} &=& c_4(S_c) \exp[c_5(S_c) z] \, ,
\eea
for which we use the coefficients $c_4$ and $c_5$ from Table 3 in \citet{Yahya_2015}.
This is then inserted in \autoref{eq:coefficients} to relate the number counts fluctuation to the underlying dark matter fluctuation $\delta_N$.

Magnification and evolution bias are then evaluated using \autoref{eq:ngal_hi} following \citep{Maartens_2021}:
\bea\label{eq:magbias_higal}
s^{\hi,gal} = -\frac{5}{2}\frac{\partial\log N_g^{\hi}}{\partial\log S_c} \, ,
\eea
\bea\label{eq:evobias_higal}
b_e^{\hi,gal} = \frac{\partial\log N_g^{\hi}}{\partial\log a} \, .
\eea



For our second radio tracer, radio continuum galaxies parameters are instead found in Table 3 in \citet{2020PASA...37....7S} for the \textit{Wide Band 1 Survey}. This lists the galaxy number density, clustering bias and magnification bias at different redshift bins. To obtain the evolution bias we compute a numerical derivative of the resulting number density with respect to redshift, similarly to \autoref{eq:evobias_higal}. 

Our third radio tracer is \hi\ intensity mapping. \hi\ intensity mapping differs to standard galaxy surveys as the method does not measure individual objects, but rather the total $21$ cm emission in each voxel (i.e. a three-dimensional pixel given by telescope beam and frequency channel width). The \hi\ brightness temperature at redshift $z$ and in direction $\boldsymbol{n}$ is proportional to the observed number of \hi\ emitters per redshift per solid angle $N_{\rm{\hi}}$:
\bea\label{eq:t_hi_def}
T_{\hi}(z,\boldsymbol{n}) \propto \frac{N_{\rm{\hi}}(z,\boldsymbol{n})}{D_A^2(z,\boldsymbol{n})} \, ,
\eea
where $D_A$ is the area distance.
This is poorly constrained by current observations and we use the fit \citet{MeerKLASS}:
\bea\label{eq:t_hi_fit}
T_{\hi} = 0.055919 + 0.23242 z - 0.024136 z^2 \, .
\eea
However, \autoref{eq:t_hi_def} implies conservation of surface brightness, yielding no magnification bias. In fact, for IM \autoref{eq:coefficients} is slightly different \citep{2013PhRvD..87f4026H} but one can recover it using the expression in Redshift Space setting
\bea\label{eq:s_hi}
s^{\hi} = \frac{2}{5} \, .
\eea
To calculate the evolution bias for intensity mapping, we follow \citet{Maartens_2021}, noticing that \autoref{eq:t_hi_def} implies that at the background level we have
\bea
T_{\rm{\hi}} \propto \frac{n_{\rm{\hi}}r^2H^{-1}}{a^2r^2} \propto \frac{(1+z)^2}{H}n_{\rm{\hi}} \, ,
\eea
where $n_{\rm{\hi}}$ is the comoving number density of \hi\ emitters in the source rest-frame. Thus we can write \citep{2018MNRAS.479.3490F,Jolicoeur_2021}:
\bea\label{eq:be_HI}
b_e^{\hi} &=& -\frac{\partial\log n^{\hi}}{\partial\log(1+z)} \nn\\
&& = -\frac{\partial\log T_{\hi}}{\partial\log(1+z)}-\frac{\partial\log H}{\partial\log(1+z)}+2 \, .
\eea
We then use \autoref{eq:t_hi_fit} and assume a $\Lambda$CDM universe for $H(z)$ to compute the evolution bias of \hi\ intensity mapping. These parameters are all inserted in \autoref{eq:coefficients} through the amplitudes of the corrections terms such as $A_L$ and $A_D$, which are expressed fully in Appendix \ref{sec:amplitudes}.

\subsection{GWs}

To model the population of GW sources we follow the prescriptions already provided in \citet{2025MNRAS.537.1912Z}. 

In this work we focus on GWs emitted by a population of binary black holes (BBHs). We do not focus on binary neutron stars (BNs) or neutron star-black hole (NSBH) mergers for simplicity. Accounting for them would require separate modelling of the intrinsic merger rate and chirp mass distribution for each, and consequently of their related magnification and evolution bias, together with assuming a form of their clustering bias. Furthermore, as the forecasted number of detections are significantly higher for BBHs with ET \citep{Punturo_2010, Iacovelli:2022bbs, Branchesi:2023mws}, we chose to only consider these events.
We assume these to follow a standard Madau-Dickinson rate from \citet{2021PhRvD.104d3507Y, Madau_2014}:
\bea \label{eq:madau}
R_{GW}(z) = R_0\frac{(1+z)^{2.7}}{1+(\frac{1+z}{2.9})^{5.6}} \, ,\
\eea
with $R_0$ providing the merger rate at $z=0$, given by \citet{GWTC2,2025PhRvD.112h4080A} as $R_0=23.9$ Gpc$^{-3}$yr$^{-1}$. The values used in \autoref{eq:madau} are not fully known and in this paper we fix them to follow a regular Madau-Dickinson model. This assumption relies on the fact that astrophysical BBHs are the results of stellar processes, and would therefore follow a rate akin to the stellar formation rate. We can then set the amplitude $R_0$ to match LVK observations, and when 3G detectors will come online we will have better constraints on the rest of the parameters. Modifications to the merger rate will result in different values of the magnification and evolution biases, thus impacting the angular power spectrum of GWs. Moreover, it will yield a new estimate of the shot noise.

The number density of observed GW sources follows previous studies \citep{Zazzera_2024,Oguri_2018} as:
\begin{equation}\label{eq:GW_numdens}
    n_{\rm GW}(z) = \tau\frac{R_{GW}(z)}{1+z}\int \d \mathcal{M}\ \phi(\mathcal{M})\,S(\rho_{th};\mathcal{M},z) \,, 
\end{equation}
where $\tau$ is the observation time of the detector, $R_{GW}(z)$ is the intrinsic merger rate in \autoref{eq:madau}, $\phi(\mathcal{M})$ is the chirp mass $\mathcal{M}$ distribution, with
\bea \label{eq:chirp_def}
\mathcal{M} \equiv \frac{(m_1m_2)^{3/5}}{(m_1+m_2)^{1/5}} \, .\
\eea
The survival function $S(\rho_{th};\mathcal{M},z)$ simply imposes whether an event of chirp mass $\mathcal{M}$ at redshift $z$ is detected assuming a certain SNR threshold $\rho_{th}$, and contains the power spectral density of the experiment, thus making each bias detector-dependent. For an O$5$-like experiment \citep{O5} we assume a detector comprising of the advanced Laser Interferometer Gravitational-Wave Observatory (LIGO) \citep{AdvLIGO}, Virgo \citep{AdvVirgo} and the Kamioka Gravitational Wave Detector (KAGRA) \citep{KAGRA:2013rdx}, whereas for ET we assume an L-shape detector with $10$km arms\footnote{The corresponding sensitivity curve for ET is found at \href{https://apps.et-gw.eu/tds/?content=3&r=18213}{https://apps.et-gw.eu/tds/?content=3\&r=18213}}\citep{BlueBook}.
An interested reader can find a full explanation of the survival function $S(\rho_{th};\mathcal{M},z)$ in \citet{Zazzera_2024, Oguri_2018, Finn_1996}.

Following \citet{Zazzera_2024}, the magnification and evolution biases for each GW experiment are defined as:
\bea
 s^{GW} &\equiv& -\frac{1}{5}\frac{\partial\log n_{GW}}{\partial\log\rho_{th}}\bigg|_{a} \, ,\\
 b_e^{GW} &\equiv& \frac{\partial\log n_{GW}}{\partial\log a}\bigg|_{\rho_{th}} \, ,
\eea
where $\rho_{th}$ is the signal-to-noise ratio (SNR) threshold of detection, which we set to $\rho_{th}=8$. This is used instead of the flux, commonly used in the galaxy counterpart.

Finally, the redshift dependence of the clustering bias is modelled as in \citet{2025MNRAS.537.1912Z} as:
\bea\label{eq:gw_clusbias}
b^{GW} = B (1+z)^\alpha \, .
\eea
where $B$ and $\alpha$ are now key parameters of interest to be constrained with our cross-correlation observables. One should note that such a simple linear model was found to suffice in describing the bias of GW events using simulations \citep{Libanore_2021}.
\autoref{eq:gw_clusbias} is then cast into \autoref{eq:coefficients} to link the observed number density contrast to the underlying dark matter fluctuations.

\section{Fisher formalism}\label{sec:fisher}
The construction of the Fisher matrices employed in this paper follows Section 4 of \citet{2025MNRAS.537.1912Z}. It hold for a GW surveys as well as any galaxy survey in the optical or in the radio. Therefore, we only report the most significant equations and changes when including \hi\ IM into the tracer's pool.

In a multi-tracer approach, using a GW survey and an \hi\ survey, we can represent schematically the data covariance matrix as:
\bea\label{eq:multitracer_cov}
\Gamma_\ell (z_i,z_j) &=& \begin{bmatrix}
\Gamma_{\ell,ij}^{\hi,\hi} & \Gamma_{\ell,ij}^{\hi,\rm GW}\\
\Gamma_{\ell,ij}^{\rm GW,\hi} & \Gamma_{\ell,ij}^{\rm GW,GW} \
\end{bmatrix} \, ,
\eea
The data covariance is defined as:
\bea\label{eq:gamma}
\Gamma^{ij}_{\ell} = C_{\ell}^{ij} + \mathcal{N}^{ij} \, ,
\eea
where $C_{\ell}^{ij}$ is the angular power spectrum of bins $i,j$ (\autoref{eq:cl_gen}), and $\mathcal{N}^{ij}$ is the related noise power spectrum, generally independent of the multipole $\ell$.
The noise angular power spectrum for a discreet tracer is dominated by the shot noise:
\bea\label{eq:shot_noise}
\mathcal{N}^A_{i} = \frac1{N^A_i} \, ,
\eea
where $N^A$ is the number of objects (galaxies or GWs) per steradian in the $i$-th bin
\bea\label{eq:shot_noise2}
N^A_i = \int \d z \ n^A(z) W(z,z_i;\Delta z_i,\sigma_i^z) \, .
\eea
Here, $ n^A$ is the comoving number density of objects as function of redshift, $W$ the window function centred at $z_i$, bin size $\Delta z_i$, and redshift scatter $\sigma^z_i = \sigma_0 (1+z)$. 

Whilst the above is applied for \hi\ galaxies and continuum, for \hi\ intensity mapping the shot noise is negligible. The relevant noise term is the instrumental variance in a \emph{voxel}. The noise contribution to the angular power spectrum is:
\bea\label{eq:noise_hi}
\mathcal{N}^{\rm instr} = T_{\rm sys}^2\frac{\Omega_{\rm surv}}{2N_{\rm dish}\tau\Delta_\nu} \, ,
\eea
where $T_{\rm sys}$ is the system temperature, $\Omega_{\rm surv}$ the sky coverage, $N_{\rm dish}$ the number of dishes employed, $\tau$ the integration time and $\Delta_\nu$ the bandwidth of each frequency channel. 

The window function $W$ to define the binning is then given by a combination of error functions following \citep{Ma_2006,Viljoen_2021}:
\bea\label{eq:window}
W(z,z_i;\Delta z_i,\sigma_i^z)  = \frac1{2} \left[ \erf\left(\frac{z_i+\delta z_i -z}{\sqrt{2}\sigma_i^z} \right) -\erf\left(\frac{z_i-\delta z_i -z}{\sqrt{2}\sigma_i^z} \right) 
\right] \, .
\eea
For galaxies and GWs $\sigma_i^z$ is large enough to produce a Gaussian-like window; however, in the case of \hi\ intensity mapping, redshift uncertainty is much lower, thus yielding a smoothed top-hat window function.

\begin{table}
    \centering
    \begin{tabular}{|c|c|c|c|c|c|c|}   
    \hline
    Survey & $z$-range & $\Delta z$ & $\sigma_z$ & $A_{survey}$ \\
    \hline
    O5 -like & $[0-1.4]$ & $0.4$ & $0.2$ & $4\pi$ \\
    ET -like & $[0-3.0]$ & $0.7$ & $0.1$ & $4\pi$ \\
    \hline
    SKA1-MID & $[0.35-3.0]$ & $0.1$ & $0.005$ & $20000 {\rm deg}^2$ \\
    SKAO \hi\ gal & $[0-0.5]$ & $0.1$ & $0.001(1+z)$ & $5000 {\rm deg}^2$ \\
    SKAO2 \hi\ gal & $[0.1-2.0]$& $0.1$ & $0.02(1+z)$ & $20000 {\rm deg}^2$ \\
    Radio continuum & $[0-3.0]$ & $3.0$ & $0.02(1+z)$ & $20000 {\rm deg}^2$ \\
    \hline
    \end{tabular}
    \caption{Summary of specifications of the surveys considered. We display the redshift range, size of $z$ bins used, i.e. $\Delta z$, the redshift scatter $\sigma_z$ and the area of the sky sampled.}
    \label{tab:sumspecs}
\end{table}

A list of the values of $\sigma_z$ is found in \autoref{tab:sumspecs} for each survey.
In the case of cross-correlating different tracers we set, for simplicity, the corresponding shot noise to zero, i.e., $\mathcal{N}^{ij}=\mathcal{N}^{i}\delta^{ij}$. Shot-noise comes from the correlation function at the same object. We do expect that some objects may overlap between a GW and galaxy catalog. One can model the cross-shot noise as proportional to the number density from the overlap in halo mass range of the two tracers weighted by the number densities of each tracer in consideration. Therefore we expect the overlap to be small, i.e., low cross-shot noise. For more details we refer to Appendix A of \citet{2020JCAP...09..054V}.
We understand that the results obtained will therefore be slightly optimistic, although in the context of a Fisher analysis we consider this acceptable.

We then multiply the \hi\ intensity mapping angular power spectrum with a beam to account for the angular resolution of the experiment
\bea
B(\theta_{\rm fwhm}) = \exp\left[-\frac{\theta^2_{\rm fwhm}}{16\log 2}\right] \, ,
\eea
with $\theta_{\rm fwhm}=1.22\frac{\lambda}{D}$ and $D$ being the antenna aperture, which for the SKAO is $D=15$m.

Finally, to account for sky localisation uncertainty we apply to the angular power spectra of GWs a beam, i.e. $B_\ell C_\ell^{GW}$. We assume that to first order we can model this as Gaussian:
\bea\label{eq:beam}
B_\ell = \exp\left(-\frac{\ell(\ell+1)}{16\ln 2}\theta^2_{\rm res}\right) \, ,
\eea
where we set the GWs resolution $\theta_{\rm res}$ to $20\degree$ for an O5-like survey \citep{Howell_2017}, and $5\degree$ for ET\citep{Libanore_2022}, consistent with distribution of localisation of 3G data \citep{Sathyaprakash_2012, Punturo_2010}. This effectively reduces the signal at smaller scales due to the limiting resolution of the detector.

The Fisher matrix is then constructed following \citet{Abramo_2022}:
\be\label{eq:fisher_entries}
F^{\mu\nu}_{\bar{\ell}}=\frac{f_{sky}}{2}\displaystyle\sum_{\ell\in\bar{\ell}}(2\ell+1)\text{Tr} \left\{ \frac{\partial\Gamma_{\ell}^{\imath\jmath}}{\partial\vartheta^{\mu}}\left[\Gamma^{\jmath\imath\prime}_{\ell}\right]^{-1}\frac{\partial\Gamma_{\ell}^{\imath\prime\jmath\prime}}{\partial\vartheta^{\nu}}\left[\Gamma^{\jmath\prime\imath}_{\ell}\right]^{-1}\right\} ,
\ee
where $\theta$ is a parameter vector and $f_{\rm sky}$ is the fraction of sky observed. For the purposes of this study, for GW detectors we approximate $f_{\rm sky}=1$. In reality, the source location relative to the GW detector network will impact its localisation. 
The sky area for galaxy surveys is found in \autoref{tab:sumspecs}.

Finally, in our Fisher matrix analyses we will always consider, and marginalise over, the standard cosmological parameters as a core set:
\bea\label{eq:cosmo_params}
\vartheta_c = \{H_0, \Omega_b,\Omega_{cdm},A_s,n_s\}\, ,
\eea
set to the fiducial values given by Planck 2018 results \citep{Planck_2018}.

\section{Clustering bias}\label{sec:clusbias}
Our first analysis is to test the synergies between GW experiments and different radio surveys in constraining the amplitude of the clustering bias of GWs from \autoref{eq:gw_clusbias}. We use a modified version of the code \texttt{CAMB} presented in \citet{2023JCAP...08..050F} in order to compute the angular power spectrum of GWs in luminosity distance space. In this first analysis we include the magnification and evolution biases to compute the $C_\ell$, however we keep them as fixed parameters to focus our attention solely on the clustering bias. Forecasts on $s^{GW}$ and $b_e^{GW}$ will follow in \autoref{sec:biases}. 

Using \autoref{eq:shot_noise} to \autoref{eq:fisher_entries} we construct Fisher matrices for each combination of surveys using the set of parameters
\bea\label{eq:params_with_bias}
\vartheta = \{ B,\alpha, \vartheta_c\}
\eea
where $B$ and $\alpha$ are defined in \autoref{eq:gw_clusbias} and $\vartheta_c$ represents the cosmological parameters and is expressed in \autoref{eq:cosmo_params}. Thus we are assuming that all parameters related to \hi\ galaxies or \hi\ intensity mapping (such as clustering, magnification and evolution bias, or \hi\ brightness temperature) are known and fixed, effectively not marginalising over them. This is similar to the forecasts described in \citet{2025MNRAS.537.1912Z}. 

\begin{figure}
    \centering
    \includegraphics[width=\linewidth]{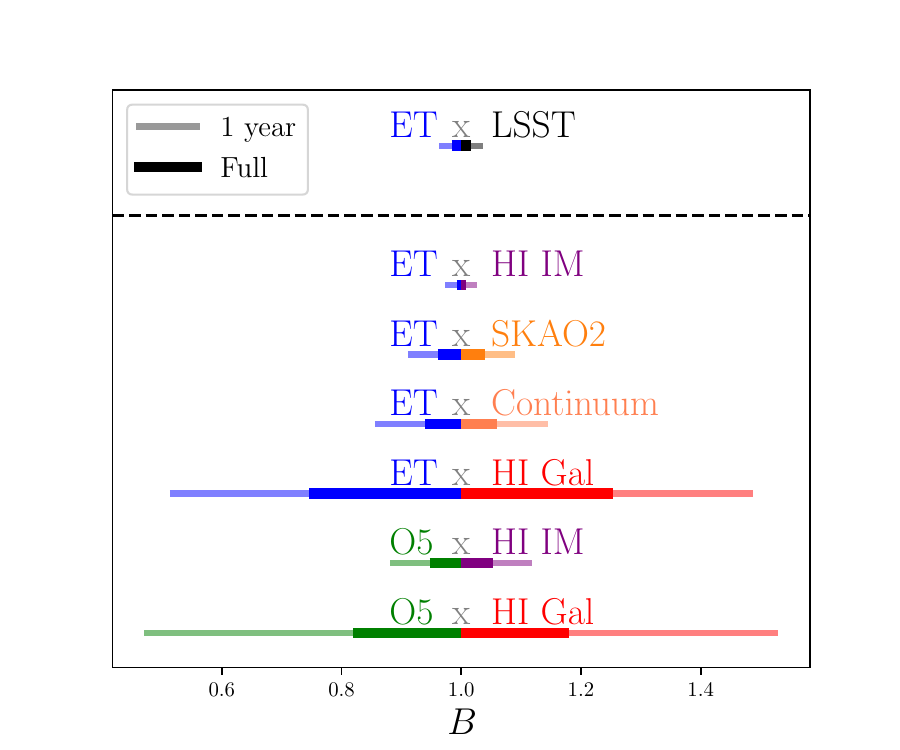}
    \caption{Measurement of the amplitude of the clustering bias of GW sources ($B$) by cross-correlating different pairs of GWs detectors and radio surveys. Faint lines show results from only $1$ year of observation, whilst bold lines include the (predicted) full length of the experiment. In the case of an ET-like detector we only adopt $5$ years of observation to show the potential with simply half their predicted run time. On top, above the dassh line, we add the result from \citet{2025MNRAS.537.1912Z} using an ET$\times$LSST-like as a mean of comparison for the results when one combines GW with radio surveys.}
    \label{fig:HI_surveys_timeline}
\end{figure}

In \autoref{fig:HI_surveys_timeline} we show the forecasts on the amplitude of the clustering bias of GWs, i.e. $B$ from \autoref{eq:gw_clusbias}, using cross-correlations between either LVK Observing run $5$ or ET, and  different \hi\ surveys. In particular, we show the $1\sigma$ uncertainty on $B$ using thin, fainter lines indicating only one year of observations, and bolder ones for results assuming a full-length survey. We find that an \hi\ galaxy survey from SKAO does not yield significant results when correlated with the next LVK Observing run or with ET. This is likely to be because of the low redshifts probed by these galaxies ($z<0.5)$, thus not having many bins which would overlap with the ones probed by GWs. Therefore, few bins would actually shave a signal for the density-density correlation, which is the one carrying the information on the clustering bias, between the two tracers (i.e. only those for which $z<0.5$). Interestingly, significant synergy is found when using \hi\ intensity mapping, both with O$5$ and with ET. In particular, cross-correlating O$5$ with an \hi\ intensity mapping survey such as the \textit{Wide Band 1 Survey} would result in $\sim 10\%$ error on $B$, whilst using data from ET would shrink the forecasted error to below $2\%$ assuming $5$ years of observations from ET and $10$k hours for SKA1-MID. 

This is by far the best result between the different cross-correlations, as the two surveys cover the same redshift range and SKA1-MID can achieve low shot noise as opposed to the other radio galaxy surveys considered. Interestingly, the results obtained with an ET$\times$\hi\ IM-like correlation are even better than those produced in \citet{2025MNRAS.537.1912Z} using an ET$\times$LSST-like one. For illustration purposes we add these at the top of \autoref{fig:HI_surveys_timeline} in black. 

\begin{figure}
    \centering
    \includegraphics[width=\linewidth]{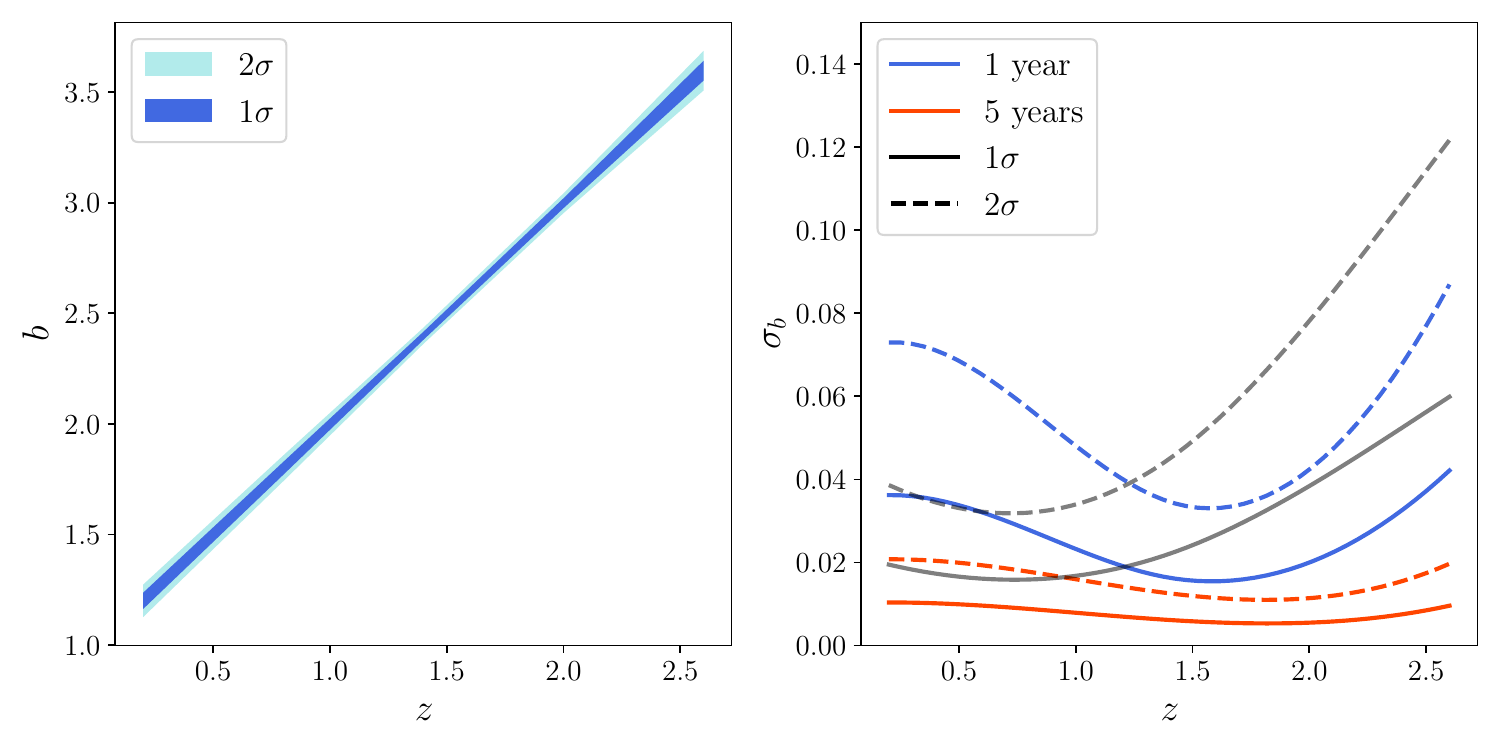}
    \caption{\textit{Left}:Projected $1$ and $2\sigma$ contour of the clustering bias $b$ of GWs as a function of redshift for a $1$ year observation period using cross-correlations of GWs from an ET-like detector and an \hi\ intensity mapping survey like SKA1-MID. \textit{Right}: Forecasted $1$ and $2\sigma$ measurement errors on the clustering bias of GWs, for both $1$ and $5$ years of observations from an ET-like detector, and $10$k hours of observation for SKA1-MID-like telescope, in blue and red lines respectively. We also add in faint grey lines the results from an ET$\times$LSST-like cross-correlation for comparison.}
    \label{fig:HI_clusbias}
\end{figure}

Considering an ET$\times$\hi\ IM-like cross-correlation, we then look at the clustering bias as a function of redshift. We do this by sampling the distribution of both $B$ and $\alpha$ and evaluating the corresponding value of the clustering bias $b$ following \autoref{eq:gw_clusbias}. Further, we compute the mean and standard deviation at every redshift value for each sample. The resulting $1$ and $2\sigma$ errors as functions of redshift are then plotted on the left hand side of \autoref{fig:HI_clusbias}. Furthermore, we plot the values of the $1$ and $2\sigma$ error on the measurement of $b$ against $z$ on the right hand panel, both for just $1$ year of observation and for $5$ years of ET data and $10$k hours of SKA1-MID. We also add, in faint grey lines, the $1$ and $2\sigma$ for an ET$\times$LSST-like cross-correlation for comparison. 

Strikingly, cross-correlating ET-like data with an \hi\ IM-like survey produces better constraints on the clustering bias of GWs across all redshifts as opposed to cross-correlating with a galaxy survey such as LSST, despite the latter covering the same $z$-range as SKA1-MID. Lower noise for the \hi\ survey increases the signal in particular at $1<z<2$, where the forecasted uncertainty reaches a minimum. This strongly reflects our modelling of the number density of GWs sources as we adopted a simple Madau-Dickinson rate (see \autoref{sec:surveys}), which in fact peaks within this range. We forecast that cross-correlating ET-like data with an \hi\ intensity mapping survey would yield competitive, if not better, results on the measurement of the clustering bias of GWs even with just one year of observations, as the $1\sigma$ error is less than $5\%$ across the whole redshift range probed.

\section{Relativistic Effects}\label{sec:relativistic}

\subsection{Measurement of GR effects}

\begin{figure*}
    \centering
    \includegraphics[width=\linewidth]{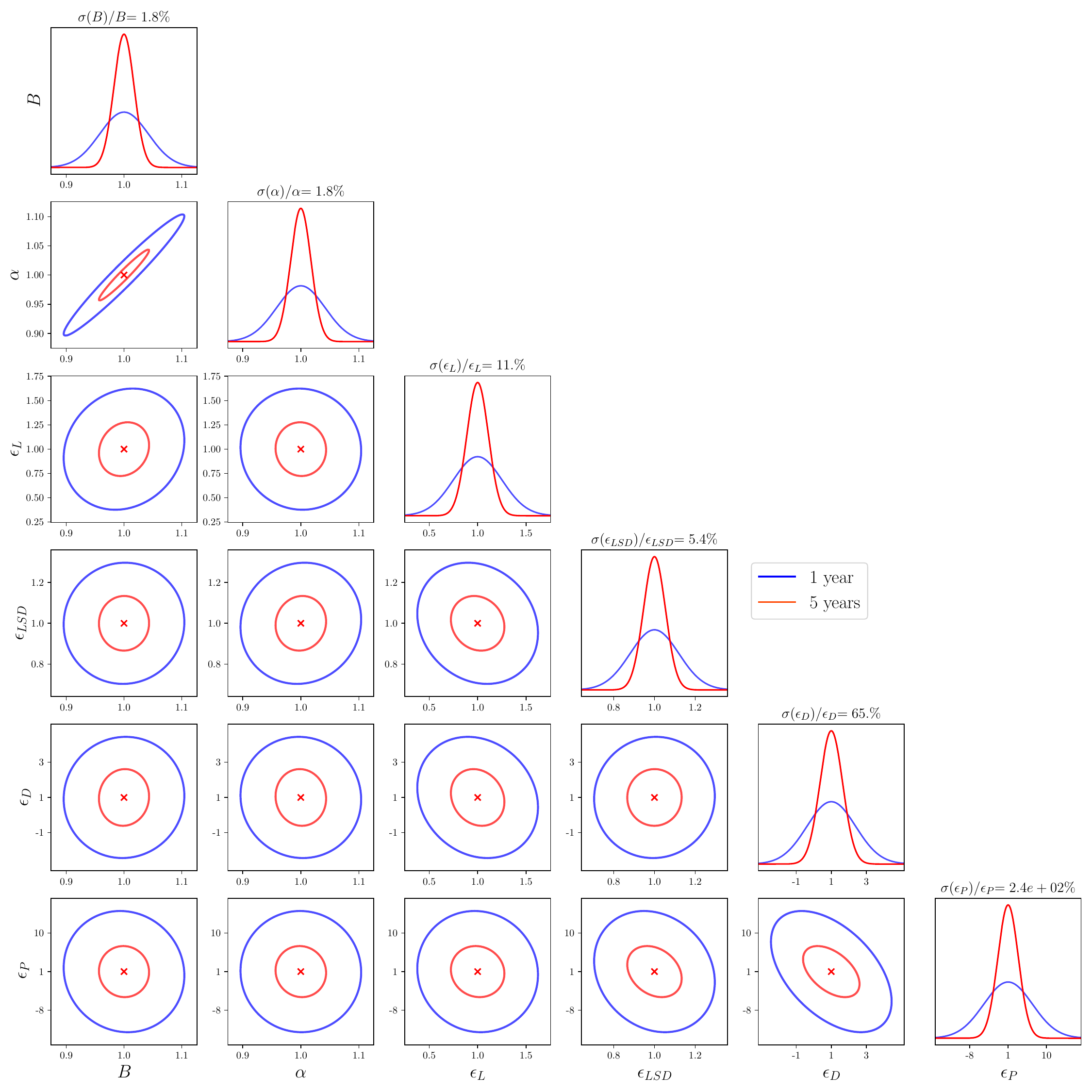}
    \caption{Predicted measurements of different relativistic corrections to the observed number counts fluctuation in a cross-correlation of ET-like GW data and \hi\ intensity mapping from an SKAO-like instrument. From top to bottom we show constraints on the two parameters of the clustering bias of GWs (i.e. $B$ and $\alpha$), then forecasted measurement of the lensing $\epsilon_L$, luminosity distance space distortions $\epsilon_{LSD}$, Doppler effect $\epsilon_D$ and finally further relativistic corrections $\epsilon_P$. We show the $1\sigma$ contours for $1$ and $5$ years worth of GWs data.}
    \label{fig:HI_epsilons}
\end{figure*}

We further explore the synergies between GWs and radio sources and turn our attention to forecasting different relativistic effects. We compute this by coupling the terms in \autoref{eq:coefficients} to a dummy amplitude $\epsilon$ which we set to $1$, implying:
\bea
\Delta=\Delta_\delta+\epsilon_{LSD}\Delta_{LSD}+\epsilon_{L}\Delta_{L}+\epsilon_{D}\Delta_{D}+\epsilon_{P}\Delta_{P}\,.
\eea
Therefore, similarly to what has been done in \citet{2025MNRAS.537.1912Z}, we forecast measurements of LSD, Lensing, Doppler and further GR corrections (mostly potentials, hence the subscript $P$) by measuring if we can constrain the respective amplitudes $\epsilon$. Note that $\epsilon_P$ is coupled to the last term in \autoref{eq:coefficients}. Considering the results of the previous section, we only forecast these measurements for an ET$\times$\hi\ IM-like cross-correlation. We present the contour ellipses in \autoref{fig:HI_epsilons}. The parameters selected for this analysis are then
\bea\label{eq:params_with_epsilons}
\theta = \{ B, \alpha, \epsilon_L, \epsilon_{LSD}, \epsilon_D, \epsilon_P, \vartheta_c\}\,,
\eea
allowing to check for degeneracies between clustering bias parameters and relativistic effects. 

In fact, one can immediately see that $B$ and $\alpha$, which build the clustering bias of GWs, are degenerate with both the lensing measurement $\epsilon_L$ and strongly with the measurement of the luminosity distance space distortions, $\epsilon_{LSD}$. 

The confidence on the measurement of the relativistic effects is greatly increased with $5$ years of GW data; for instance the error on $\epsilon_L$ shrinks of a factor of $10$ to roughly $10\%$. Notably, the signal on the LSD contribution, i.e. the radial velocity effect, is particularly strong, reaching an error of $\sim7\%$ with $5$ years-worth of data. Doppler and further relativistic corrections remain unobservable instead. 

It is interesting to compare these results with those presented with a cross-correlation of GWs from a third generation detector with photometric galaxies from an LSST-like survey \citep{2025MNRAS.537.1912Z}. Whilst the lensing appears more easily detectable through correlations with galaxies (reaching and error of $\sim3\%$), the error on $\epsilon_{LSD}$ is improved by a few percent with cross-correlations with intensity mapping, going below the $10\%$ error from the previous study. In the case of intensity mapping, lensing is suppressed as hinted in \autoref{sec:surveys}. Therefore, the $C_{\ell}$ in \autoref{eq:multitracer_cov} will include lensing-lensing correlations only in the lower right quadrant, i.e. the auto-correlation of GWs. In fact, as $\Delta_{L}^{\hi}=0$, both auto-correlation of \hi\ lensing and cross-correlation \hi\ lensing $\times$ GW lensing will yield zero. Therefore, the signal on the lensing correction, $\epsilon_L$ is greatly reduced. However, this implies then that the contribution to the number count fluctuation in \autoref{eq:coefficients} given by the LSD term is much more significant, yielding in fact a stronger signal than the cross-correlation with galaxies.

\subsection{Magnification and evolution bias}\label{sec:biases}

\begin{figure}
    \centering
    \includegraphics[width=\linewidth]{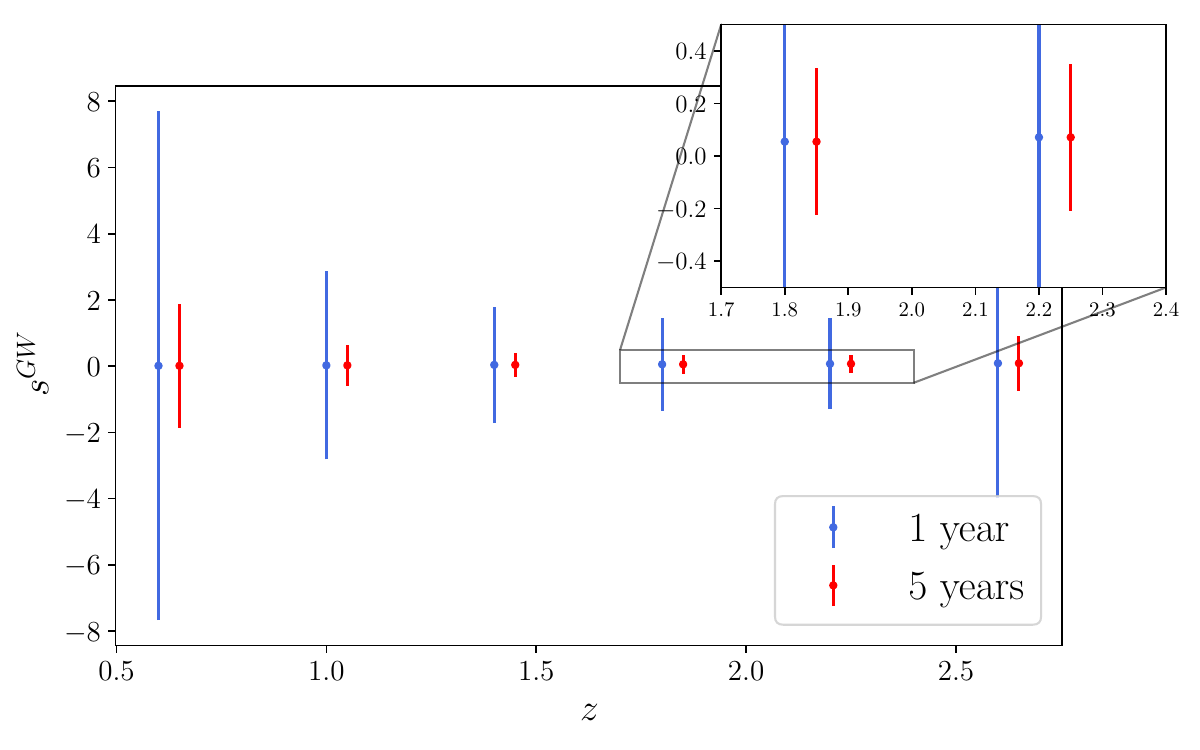}
    \includegraphics[width=\linewidth]{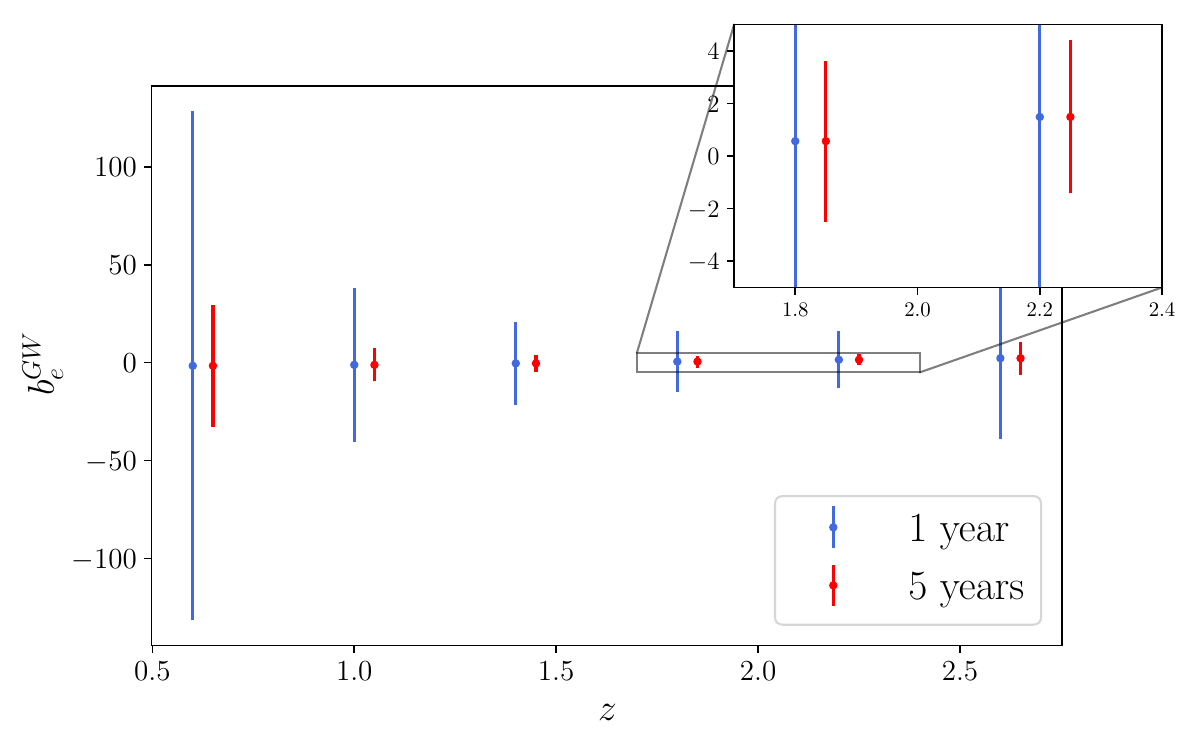}
    \caption{Constraints on the magnification (\textit{top}) and evolution (\textit{bottom}) biases of GWs for an ET-like experiment, obtained by cross-correlating GWs with an IM survey such as SKA1-MID.}
    \label{fig:HIxGW_s_be_biases}
\end{figure}

We also produce forecasts on the magnification and evolution biases of GWs, in a similar fashion to \citet{2025MNRAS.537.1912Z}, that is, constraining the values of $s^{GW}$ and $b_e^{GW}$ at different redshift bins. Thus, we construct a Fisher matrix with parameters
\bea\label{eq:params_with_magevobias}
\vartheta = \{ B, \alpha, s^{GW}(z_k), b_e^{GW}(z_k), \vartheta_c\}\,,
\eea
where $z_k$ runs over the $z$-bins. However, whilst ET's redshift range starts close to the observer, \hi\ intensity mapping covers the range $z \in [0.35,3.0]$. Thus, we show only the constraints on the biases at redshifts covered by both surveys. Similarly to the previous section, we only show results with the combination of surveys that provided the best results in \autoref{sec:clusbias}, i.e. ET$\times$\hi\ IM-like. We display the results in \autoref{fig:HIxGW_s_be_biases}, where on the top panel we show the forecasted measure of an ET-like $s^{GW}$ and on the bottom panel the corresponding $b_e^{GW}$. 

The results are competitive with the ones displayed in our previous study using an ET$\times$LSST-like cross-correlation, especially in the redshift range $1.5<z<2.5$, where shot noise of GWs is greatly reduced due to the higher number of sources in the Madau-Dickinson rate used. 

\begin{figure*}
    \centering
    \includegraphics[width=\linewidth]{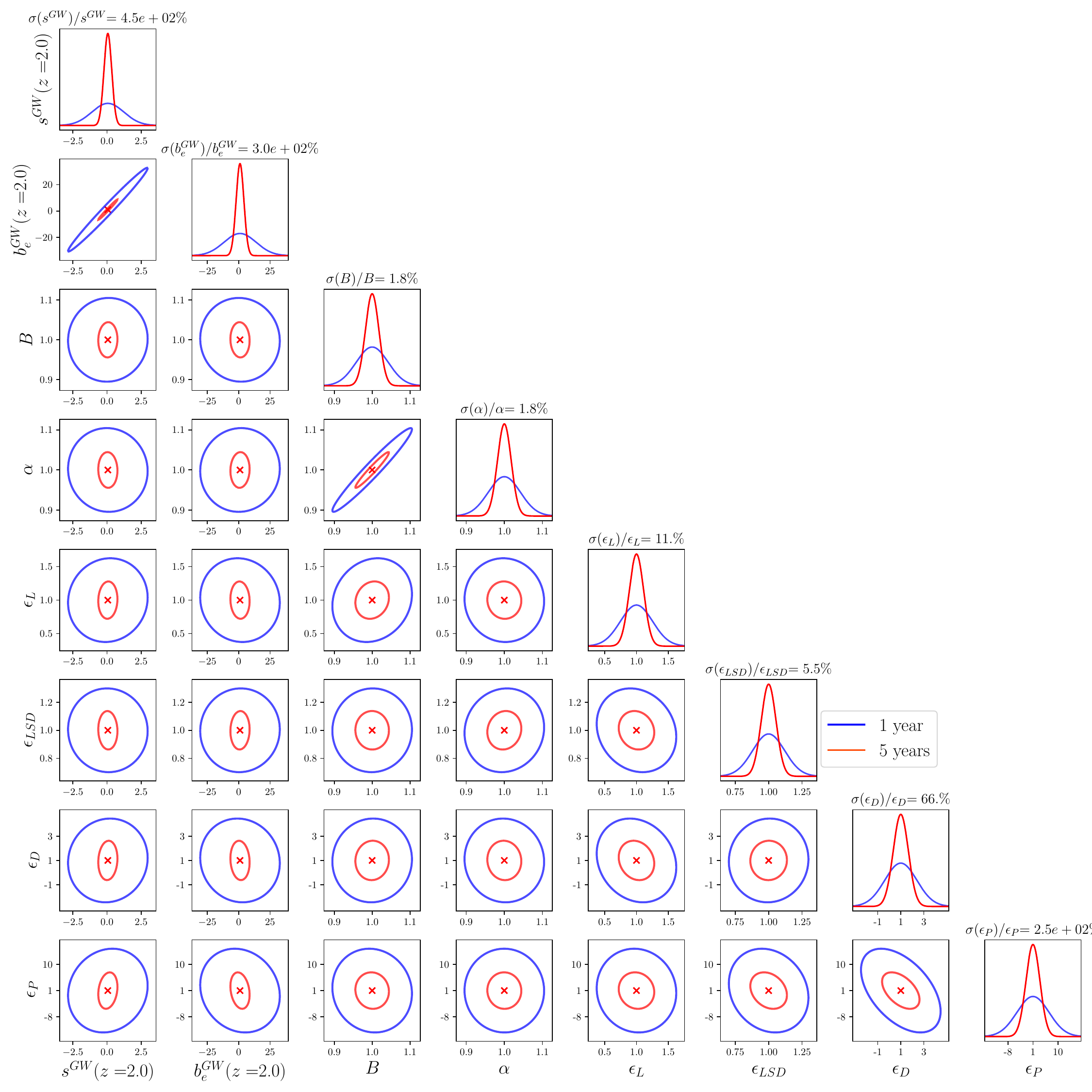}
    \caption{Constraints on magnification and evolution biases of GW sources, $s^{GW}$ and $b_e^{GW}$, at $z=2$, together with constraints on clustering bias parameters $B$ and $\alpha$, and the measurements of the relativistic effects $\epsilon_L,\epsilon_{LSD},\epsilon_D,\epsilon_P$. These are obtained through a cross-correlation of GWs from an ET-like detector and IM survey such as SKA1-MID.}
    \label{fig:triangle_ALL_GWxHI}
\end{figure*}

We then check for a possible degeneracy between $s^{GW}$ and $b_e^{GW}$ and the other parameters $b,\alpha,\epsilon_L,\epsilon_{LSD},\epsilon_D,\epsilon_P$. As an example, we show the forecasts on these parameters (looking at $s^{GW}(z=2.0)$ and $b_e^{GW}(z=2.0)$) in \autoref{fig:triangle_ALL_GWxHI}, showing no strong degeneracy between the biases and the other parameters.

\section{Extension with galaxies}\label{sec:triple}

\begin{figure}
    \centering
    \includegraphics[width=\linewidth]{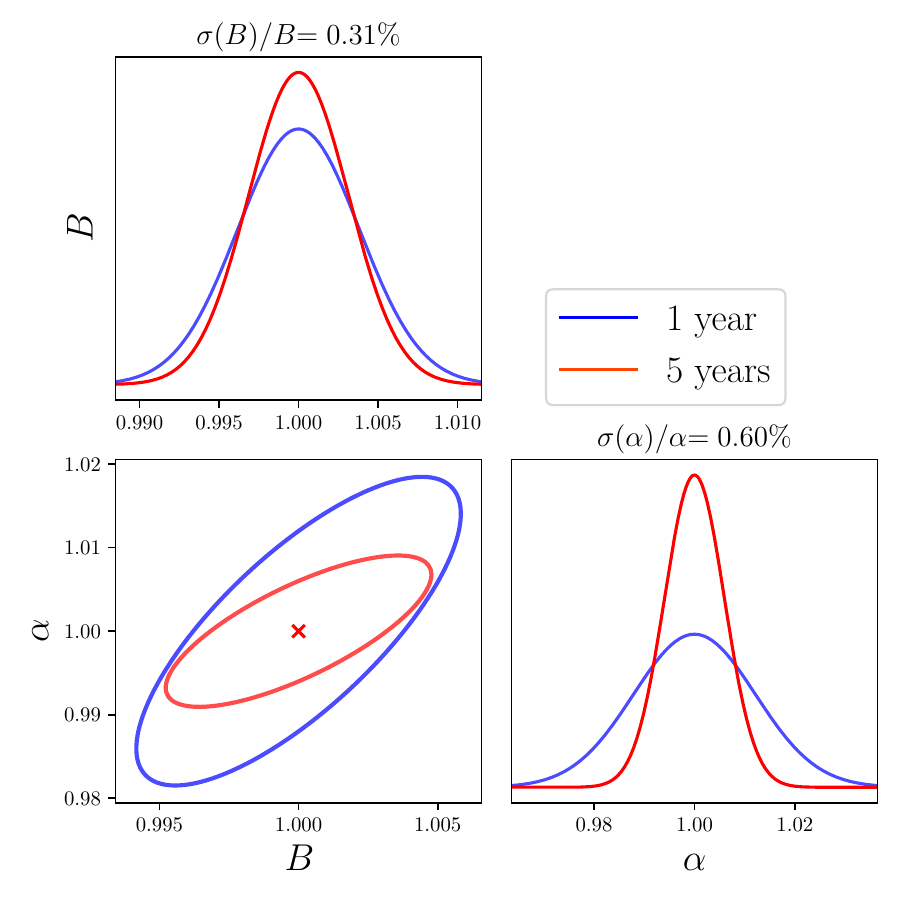}
    \caption{Constraints on clustering parameters $B$ and $\alpha$ using a three tracer correlation with a \hi\ intensity mapping survey like with SKAO1-MID, an LSST-like photometric galaxy survey, and GWs from an ET-like experiment.}
    \label{fig:triple_balpha}
\end{figure}

Having explored synergies between future GW experiments and radio surveys, we now extend this work to include all possible tracers by adding optical galaxy surveys. In particular, we can construct a cross-correlation of ET-like data, \hi\ IM data from an SKA1-MID-like survey, and photometric galaxies from an LSST-like survey. Details of the latter can be found in \citet{2025MNRAS.537.1912Z}. Effectively, this constructs all possible cross-correlations between the three tracers: \hi$\times$\hi, \hi$\times$GWs, \hi$\times$Gal, GW$\times$Gal etc... and produces a data covariance matrix of the form:

\bea\label{eq:triple_cov}
\Gamma_\ell (z_i,z_j) &=& \begin{bmatrix}
\Gamma_{\ell,ij}^{\hi,\hi} & \Gamma_{\ell,ij}^{\hi,\rm Gal} & \Gamma_{\ell,ij}^{\hi,\rm GW}\\
\Gamma_{\ell,ij}^{\rm Gal,\hi} & \Gamma_{\ell,ij}^{\rm Gal,Gal} & \Gamma_{\ell,ij}^{\rm Gal,GW}\\
\Gamma_{\ell,ij}^{\rm GW,\hi} & \Gamma_{\ell,ij}^{\rm GW,Gal} & \Gamma_{\ell,ij}^{\rm GW,GW} \
\end{bmatrix} \, .
\eea
We can then produce forecasts over the full list of parameters directly
\be
\vartheta = \{ B,\alpha,s^{GW}(z_k),b_e^{GW}(z_k),\epsilon_L,\epsilon_{LSD},\epsilon_D,\epsilon_P,\vartheta_c\}\,.
\ee
Firstly, we show the results for the parameters of the clustering bias in \autoref{fig:triple_balpha}. These yield sub-percent precision on the amplitude $B$ as opposed to the $\sim2\%$ error predicted in \autoref{sec:clusbias} using only GWs and intensity mapping.

\begin{figure}
    \centering
    \includegraphics[width=\linewidth]{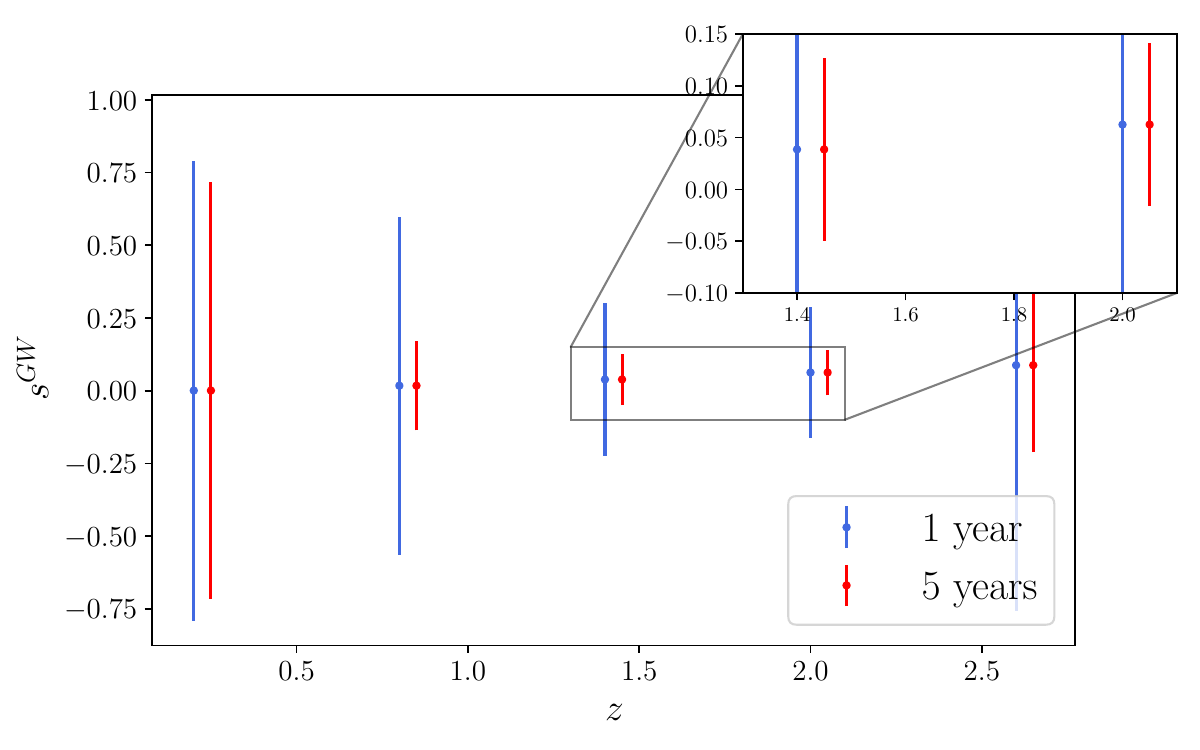}
    \includegraphics[width=\linewidth]{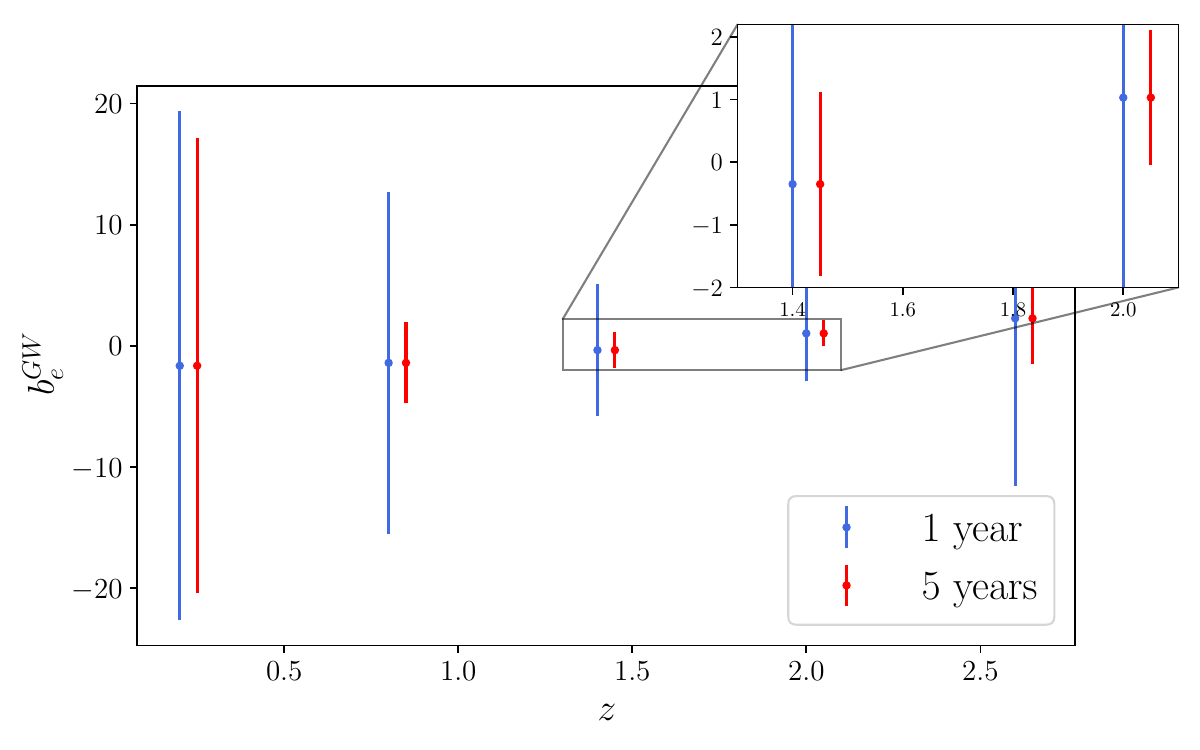}
    \caption{Constraints on the magnification (top panel) and evolution (bottom panel) biases of GW sources using a triple correlation of intensity mapping, photometric galaxies and GWs.}
    \label{fig:triple_biases}
\end{figure}

The increase in precision is also found in the constraints on both the magnification and evolution biases, $s^{GW}$ and $b_e^{GW}$, which we show in \autoref{fig:triple_biases}. 

\begin{figure*}
    \centering
    \includegraphics[width=\linewidth]{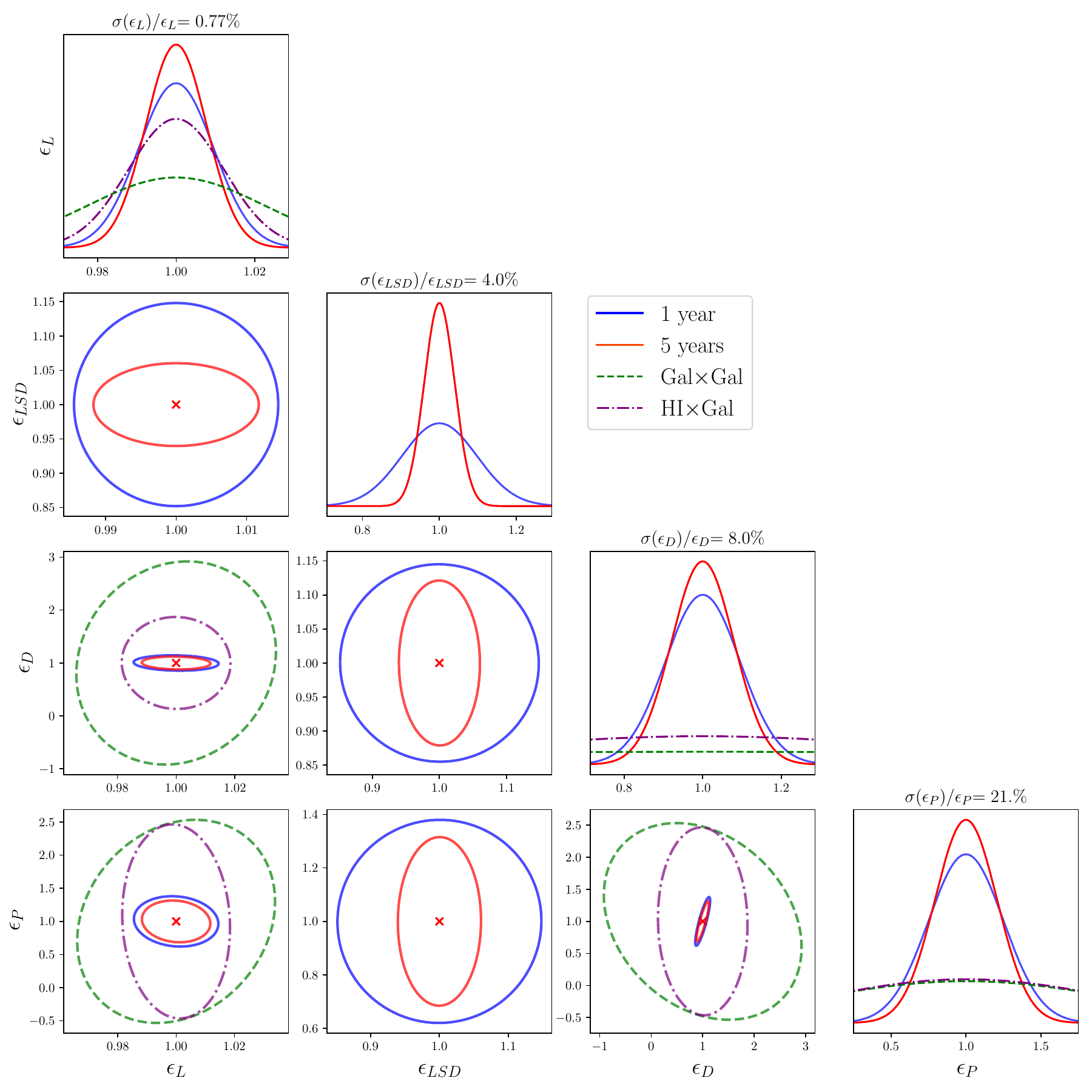}
    \caption{Constraints on the measurements of different relativistic effects by computing a cross-correlation of an intensity mapping survey, a Stage IV galaxy survey and a third-generation GW detector. We show the errors on the lensing effect $\epsilon_L$, on the Doppler term $\epsilon_D$ and on the subdominant relativistic effects $\epsilon_P$. We show these results for the triple correlation assuming $1$ year of observation (in blue) and $5$ years (in red). We then include constraints on the same parameters using only the autocorrelation of the galaxy survey (in green) and then the cross-correlation \hi\ IM$\times$Galaxies. We also show the constraints on $\epsilon_{LSD}$ only coming for the triple correlation, as only GWs carry information on the luminosity distance space distortions.}
    \label{fig:triple_allsurveys}
\end{figure*}

Finally, we also show the constraints on the measurement of different relativistic effects (Lensing, LSD, Doppler and the subdominant GR effects - see \autoref{sec:relativistic}) in \autoref{fig:triple_allsurveys}. We show the constraints given by this triple correlation of tracers using both $1$ year and $5$ years of data. To ascertain whether a particular tracer - or combination of - is dominating the information on these parameters, we also include the auto-correlation of galaxies (in green) and the cross-correlation of galaxies and intensity mapping (in purple), both assuming $5$ years of data. As both galaxies and neutral hydrogen are redshift tracers, we do not include constraints on the radial velocity distortions in luminosity distance space - i.e. the LSD - as they instead carry information on the redshift space distortions (RSDs).

Interestingly, whilst the autocorrelation of galaxies alone can yield around a $5\%$ uncertainty on the measurement of the lensing correction, including GWs brings the error down to lower than $2\%$. Adding GWs can also significantly increase the precision on other relativistic effects. In particular, the signal on the Doppler term goes from being poorly constrained to being measured with a rough uncertainty of $\sim 15\%$.

\section{Discussion and conclusions}\label{sec:conclusions}

In this paper we analysed synergies between future GW experiments and upcoming radio surveys. In particular, we presented forecasts on the clustering, magnification and evolution bias of gravitational waves from binary black hole mergers, and on the detectability of different relativistic effects. 

We established that cross-correlating GW data from a third-generation ground based detector and an intensity mapping survey such as SKA1-MID would produce high quality constraints on the clustering bias of GWs, achieving an error on the amplitude of $b^{GW}$ lower than $2\%$. Interestingly, intensity mapping showed the best synergy with GW experiments, resulting in good precision on the amplitude of $b^{GW}$ when cross-correlating with data from an O$5$-like experiment as well, achieving precision of about $5\%$ using the predicted full length of the O$5$ run. This can be explained by the large redshift range covered by an intensity mapping survey such as SKA1-MID. In fact, this would cover a range $0<z<3$, perfectly overlapping with the range covered by ET, and covering the entirety of that of O$5$. The more bins overlap and the larger the signal on the density term in \autoref{eq:coefficients} will be, which is the dominant term in the number counts fluctuation (at least until high redshifts). Additionally, the shot noise for an SKAO intensity mapping survey is predicted to be advantageous compared to other tracers such as radio galaxies, thus improving the error on the parameters considered.

Further, we also compare our forecasted results with a previous analysis cross-correlating GWs with photometric galaxies from an LSST-like survey \citep{2025MNRAS.537.1912Z}. We show that cross-correlating GWs and intensity mapping can yield better results on the clustering bias of gravitational waves, especially when considering the error on $b^{GW}$ as a function of redshift. In \autoref{fig:HI_clusbias} we show that simply cross-correlating $1$ year of GW data with an intensity mapping survey such as SKA1-MID can produce constraints similar to those obtained by cross-correlating $5$ years of GW and photometric galaxy data. Additionally, we forecast that with $5$ years of observation, an ET$\times$\hi\ correlation would yield a $2\sigma$ precision of less than $2\%$ up until $z=2.5$. 

We also produce constraints on the magnification and evolution biases of GWs. Initially, cross-correlating GWs with intensity mapping produced results comparable to the ones obtained by cross-correlating with photometric galaxies \citep{2025MNRAS.537.1912Z}. However, significantly better results are obtained with a triple correlation of tracers, i.e. GWs, intensity mapping and photometric galaxies. Using all three tracers, the uncertainty on the values of $s^{GW}$ and $b_e^{GW}$ is reduced by a factor of $10$. As these parameters are dependent on both the intrinsic merger rate of BBH mergers and on their chirp mass distribution, constraining them would provide a direct handle on constraining these intrinsic properties of BBHs.
\citet{Zazzera_2024} showed that two slightly different primary mass distribution, \textit{Power-Law + Peak} (used in this paper) and \textit{Broken Power-Law}, result in almost identical expressions of the biases, thus having a much less significant impact. Therefore, constraining the values of $s^{GW}$ and $b_e^{GW}$ would yield a consistency check with the measured merger rate of BBHs.

Using a cross-correlation of these three different tracers, we also forecast constraints on different relativistic effects, namely the lensing correction, luminosity distance space distortions, Doppler and subdominant GR effects. Significant constraints can be achieved for the lensing correction also with only the galaxy autocorrelation or the cross-correlation of photometric galaxies and intensity mapping. However, adding gravitational waves increases the precision on the measurement of these effects by a large amount. We forecast an uncertainty on the measurement of the lensing term of less than $2\%$, and the possibility of measuring the Doppler term with around $15\%$ error, thus at a $6.7\sigma$ level. Furthermore we show for the first time that the gravitational potential terms can be detected in LSS (at $3\sigma$), which are below the cosmic variance limit for two tracers. This is an important result, as it opens up the possibility of testing gravity in large scale structure studies with GWs. Different modified gravity theories would result in an alternative expression for the number counts (as investigated in e.g. \citet{2024JCAP...02..023B}), especially affecting terms such as the lensing term. Measuring the latter would then help constrain different modifications of gravity. 

To conclude, we highlight a synergy between intensity mapping and future GW experiments, particularly as cross-correlations of the two tracers would yield precise measurements of the clustering bias of gravitational waves. Furthermore, we showed that cross-correlating GW, intensity mapping and photometric galaxies altogether would produce tight measurements of the lensing effect on the number counts ($<2\%$) and other relativistic effects such as the luminosity distance space distortions and Doppler term. 

\section*{Data Availability}
The data underlying this article will be shared on reasonable request to the corresponding author.

\section*{Acknowledgements}
We are pleased to thank Roy Maartens and Anna Balaudo for useful discussion. S.Z. acknowledges support from the Perren Fund, University of London. JF acknowledges support of Funda\c{c}\~{a}o para a Ci\^{e}ncia e a Tecnologia through the Investigador FCT Contract No. 2020.02633.CEECIND/CP1631/CT0002, and the research grants UIDB/04434/2020 and UIDP/04434/2020. T.B. is supported by ERC Starting Grant \textit{SHADE} (grant no.~StG 949572) and a Royal Society University Research Fellowship (grant no.~URF$\backslash$R$\backslash$231006).

\bibliographystyle{mnras}
\bibliography{forecastbib}

@ARTICLE{2020JCAP...09..054V,
       author = {{Viljoen}, Jan-Albert and {Fonseca}, Jos{\'e} and {Maartens}, Roy},
        title = "{Constraining the growth rate by combining multiple future surveys}",
      journal = {\jcap},
         year = 2020,
        month = sep,
       volume = {2020},
       number = {9},
          eid = {054},
        pages = {054},
          doi = {10.1088/1475-7516/2020/09/054},
archivePrefix = {arXiv},
       eprint = {2007.04656},
 primaryClass = {astro-ph.CO},
       adsurl = {https://ui.adsabs.harvard.edu/abs/2020JCAP...09..054V}
}

@article{Challinor_2011,
	doi = {10.1103/physrevd.84.043516},
  
  
	year = 2011,
	month = {aug},
  
	publisher = {American Physical Society ({APS})},
  
	volume = {84},
  
	number = {4},
  
	author = {Anthony Challinor and Antony Lewis},
  
	title = {Linear power spectrum of observed source number counts},
  
	journal = {Physical Review D}
}

@article{Bonvin_2008,
	doi = {10.1103/physrevd.78.123530},
  
  
	year = 2008,
	month = {dec},
  
	publisher = {American Physical Society ({APS})},
  
	volume = {78},
  
	number = {12},
  
	author = {Camille Bonvin},
  
	title = {Effect of peculiar motion in weak lensing},
  
	journal = {Physical Review D}
}

@article{Madau_2014,
	doi = {10.1146/annurev-astro-081811-125615},
  
  
	year = 2014,
	month = {aug},
  
	publisher = {Annual Reviews},
  
	volume = {52},
  
	number = {1},
  
	pages = {415--486},
  
	author = {Piero Madau and Mark Dickinson},
  
	title = {Cosmic Star-Formation History},
  
	journal = {Annual Review of Astronomy and Astrophysics}
}

@article{Oguri_2018,
	doi = {10.1093/mnras/sty2145},

  
	year = 2018,
	month = {aug},
  
	publisher = {Oxford University Press ({OUP})},
  
	volume = {480},
  
	number = {3},
  
	pages = {3842--3855},
  
	author = {Masamune Oguri},
  
	title = {Effect of gravitational lensing on the distribution of gravitational waves from distant binary black hole mergers},
  
	journal = {Monthly Notices of the Royal Astronomical Society}
}

@article{Libanore_2021,
	doi = {10.1088/1475-7516/2021/02/035},

	year = 2021,
	month = {feb},
  
	publisher = {{IOP} Publishing},
  
	volume = {2021},
  
	number = {02},
  
	pages = {035--035},
  
	author = {S. Libanore and M. C. Artale and D. Karagiannis and M. Liguori and N. Bartolo and Y. Bouffanais and N. Giacobbo and M. Mapelli and S. Matarrese},
  
	title = {Gravitational Wave mergers as tracers of Large Scale Structures},
  
	journal = {Journal of Cosmology and Astroparticle Physics}
}

@article{LSST_2021,
	doi = {10.3847/1538-4365/abd62c},
  
	year = 2021,
	month = {mar},
  
	publisher = {American Astronomical Society},
  
	volume = {253},
  
	number = {1},
  
	pages = {31},
  
	author = {{LSST Science Collaboration}, Bela Abolfathi and David Alonso and Robert Armstrong and {\'{E}
}ric Aubourg and Humna Awan and Yadu N. Babuji and Franz Erik Bauer and Rachel Bean and George Beckett and Rahul Biswas and Joanne R. Bogart and Dominique Boutigny and Kyle Chard and James Chiang and Chuck F. Claver and Johann Cohen-Tanugi and C{\'{e}}line Combet and Andrew J. Connolly and Scott F. Daniel and Seth W. Digel and Alex Drlica-Wagner and Richard Dubois and Emmanuel Gangler and Eric Gawiser and Thomas Glanzman and Phillipe Gris and Salman Habib and Andrew P. Hearin and Katrin Heitmann and Fabio Hernandez and Ren{\'{e}}e Hlo{\v{z}}ek and Joseph Hollowed and Mustapha Ishak and {\v{Z}}eljko Ivezi{\'{c}} and Mike Jarvis and Saurabh W. Jha and Steven M. Kahn and J. Bryce Kalmbach and Heather M. Kelly and Eve Kovacs and Danila Korytov and K. Simon Krughoff and Craig S. Lage and Fran{\c{c}}ois Lanusse and Patricia Larsen and Laurent Le Guillou and Nan Li and Emily Phillips Longley and Robert H. Lupton and Rachel Mandelbaum and Yao-Yuan Mao and Phil Marshall and Joshua E. Meyers and Marc Moniez and Christopher B. Morrison and Andrei Nomerotski and Paul O'Connor and HyeYun Park and Ji Won Park and Julien Peloton and Daniel Perrefort and James Perry and St{\'{e}}phane Plaszczynski and Adrian Pope and Andrew Rasmussen and Kevin Reil and Aaron J. Roodman and Eli S. Rykoff and F. Javier S{\'{a}}nchez and Samuel J. Schmidt and Daniel Scolnic and Christopher W. Stubbs and J. Anthony Tyson and Thomas D. Uram and Antonio Villarreal and Christopher W. Walter and Matthew P. Wiesner and W. Michael Wood-Vasey and Joe Zuntz},
  
	title = {The {LSST} {DESC} {DC}2 Simulated Sky Survey},
  
	journal = {The Astrophysical Journal Supplement Series}
}

@misc{LSST_book,
      title={LSST Science Book, Version 2.0}, 
      author={{LSST Science Collaboration} and Paul A. Abell and Julius Allison and Scott F. Anderson and John R. Andrew and J. Roger P. Angel and Lee Armus and David Arnett and S. J. Asztalos and Tim S. Axelrod and Stephen Bailey and D. R. Ballantyne and Justin R. Bankert and Wayne A. Barkhouse and Jeffrey D. Barr and L. Felipe Barrientos and Aaron J. Barth and James G. Bartlett and Andrew C. Becker and Jacek Becla and Timothy C. Beers and Joseph P. Bernstein and Rahul Biswas and Michael R. Blanton and Joshua S. Bloom and John J. Bochanski and Pat Boeshaar and Kirk D. Borne and Marusa Bradac and W. N. Brandt and Carrie R. Bridge and Michael E. Brown and Robert J. Brunner and James S. Bullock and Adam J. Burgasser and James H. Burge and David L. Burke and Phillip A. Cargile and Srinivasan Chandrasekharan and George Chartas and Steven R. Chesley and You-Hua Chu and David Cinabro and Mark W. Claire and Charles F. Claver and Douglas Clowe and A. J. Connolly and Kem H. Cook and Jeff Cooke and Asantha Cooray and Kevin R. Covey and Christopher S. Culliton and Roelof de Jong and Willem H. de Vries and Victor P. Debattista and Francisco Delgado and Ian P. Dell'Antonio and Saurav Dhital and Rosanne Di Stefano and Mark Dickinson and Benjamin Dilday and S. G. Djorgovski and Gregory Dobler and Ciro Donalek and Gregory Dubois-Felsmann and Josef Durech and Ardis Eliasdottir and Michael Eracleous and Laurent Eyer and Emilio E. Falco and Xiaohui Fan and Christopher D. Fassnacht and Harry C. Ferguson and Yanga R. Fernandez and Brian D. Fields and Douglas Finkbeiner and Eduardo E. Figueroa and Derek B. Fox and Harold Francke and James S. Frank and Josh Frieman and Sebastien Fromenteau and Muhammad Furqan and Gaspar Galaz and A. Gal-Yam and Peter Garnavich and Eric Gawiser and John Geary and Perry Gee and Robert R. Gibson and Kirk Gilmore and Emily A. Grace and Richard F. Green and William J. Gressler and Carl J. Grillmair and Salman Habib and J. S. Haggerty and Mario Hamuy and Alan W. Harris and Suzanne L. Hawley and Alan F. Heavens and Leslie Hebb and Todd J. Henry and Edward Hileman and Eric J. Hilton and Keri Hoadley and J. B. Holberg and Matt J. Holman and Steve B. Howell and Leopoldo Infante and Zeljko Ivezic and Suzanne H. Jacoby and Bhuvnesh Jain and R and Jedicke and M. James Jee and J. Garrett Jernigan and Saurabh W. Jha and Kathryn V. Johnston and R. Lynne Jones and Mario Juric and Mikko Kaasalainen and Styliani and Kafka and Steven M. Kahn and Nathan A. Kaib and Jason Kalirai and Jeff Kantor and Mansi M. Kasliwal and Charles R. Keeton and Richard Kessler and Zoran Knezevic and Adam Kowalski and Victor L. Krabbendam and K. Simon Krughoff and Shrinivas Kulkarni and Stephen Kuhlman and Mark Lacy and Sebastien Lepine and Ming Liang and Amy Lien and Paulina Lira and Knox S. Long and Suzanne Lorenz and Jennifer M. Lotz and R. H. Lupton and Julie Lutz and Lucas M. Macri and Ashish A. Mahabal and Rachel Mandelbaum and Phil Marshall and Morgan May and Peregrine M. McGehee and Brian T. Meadows and Alan Meert and Andrea Milani and Christopher J. Miller and Michelle Miller and David Mills and Dante Minniti and David Monet and Anjum S. Mukadam and Ehud Nakar and Douglas R. Neill and Jeffrey A. Newman and Sergei Nikolaev and Martin Nordby and Paul O'Connor and Masamune Oguri and John Oliver and Scot S. Olivier and Julia K. Olsen and Knut Olsen and Edward W. Olszewski and Hakeem Oluseyi and Nelson D. Padilla and Alex Parker and Joshua Pepper and John R. Peterson and Catherine Petry and Philip A. Pinto and James L. Pizagno and Bogdan Popescu and Andrej Prsa and Veljko Radcka and M. Jordan Raddick and Andrew Rasmussen and Arne Rau and Jeonghee Rho and James E. Rhoads and Gordon T. Richards and Stephen T. Ridgway and Brant E. Robertson and Rok Roskar and Abhijit Saha and Ata Sarajedini and Evan Scannapieco and Terry Schalk and Rafe Schindler and Samuel Schmidt and Sarah Schmidt and Donald P. Schneider and German Schumacher and Ryan Scranton and Jacques Sebag and Lynn G. Seppala and Ohad Shemmer and Joshua D. Simon and M. Sivertz and Howard A. Smith and J. Allyn Smith and Nathan Smith and Anna H. Spitz and Adam Stanford and Keivan G. Stassun and Jay Strader and Michael A. Strauss and Christopher W. Stubbs and Donald W. Sweeney and Alex Szalay and Paula Szkody and Masahiro Takada and Paul Thorman and David E. Trilling and Virginia Trimble and Anthony Tyson and Richard Van Berg and Daniel Vanden Berk and Jake VanderPlas and Licia Verde and Bojan Vrsnak and Lucianne M. Walkowicz and Benjamin D. Wandelt and Sheng Wang and Yun Wang and Michael Warner and Risa H. Wechsler and Andrew A. West and Oliver Wiecha and Benjamin F. Williams and Beth Willman and David Wittman and Sidney C. Wolff and W. Michael Wood-Vasey and Przemek Wozniak and Patrick Young and Andrew Zentner and Hu Zhan},
      year={2009},
      eprint={0912.0201},
      archivePrefix={arXiv},
      primaryClass={astro-ph.IM}
}

@article{Libanore_2022,
	doi = {10.1088/1475-7516/2022/02/003},
  
	year = 2022,
	month = {feb},
  
	publisher = {{IOP} Publishing},
  
	volume = {2022},
  
	number = {02},
  
	pages = {003},
  
	author = {S. Libanore and M.C. Artale and D. Karagiannis and M. Liguori and N. Bartolo and Y. Bouffanais and M. Mapelli and S. Matarrese},
  
	title = {Clustering of Gravitational Wave and Supernovae events:  a multitracer analysis  in Luminosity Distance Space},
  
	journal = {Journal of Cosmology and Astroparticle Physics}
}

@article{Scelfo_2018,
	doi = {10.1088/1475-7516/2018/09/039},
  
	year = 2018,
	month = {sep},
  
	publisher = {{IOP} Publishing},
  
	volume = {2018},
  
	number = {09},
  
	pages = {039--039},
  
	author = {Giulio Scelfo and Nicola Bellomo and Alvise Raccanelli and Sabino Matarrese and Licia Verde},
  
	title = {{GW}{\texttimes}{LSS}: chasing the progenitors of merging binary black holes},
  
	journal = {Journal of Cosmology and Astroparticle Physics}
}

@article{Scelfo_2020,
	doi = {10.1088/1475-7516/2020/10/045},
  
	year = 2020,
	month = {oct},
  
	publisher = {{IOP} Publishing},
  
	volume = {2020},
  
	number = {10},
  
	pages = {045--045},
  
	author = {Giulio Scelfo and Lumen Boco and Andrea Lapi and Matteo Viel},
  
	title = {Exploring galaxies-gravitational waves cross-correlations as an astrophysical probe},
  
	journal = {Journal of Cosmology and Astroparticle Physics}
}

@article{Scelfo_2022_2,
	doi = {10.1088/1475-7516/2023/02/010},
  
	year = 2023,
	month = {feb},
  
	publisher = {{IOP} Publishing},
  
	volume = {2023},
  
	number = {02},
  
	pages = {010},
  
	author = {Giulio Scelfo and Maria Berti and Alessandra Silvestri and Matteo Viel},
  
	title = {Testing gravity with gravitational waves {\texttimes} electromagnetic probes cross-correlations},
  
	journal = {Journal of Cosmology and Astroparticle Physics}
}

@article{Sathyaprakash_2012,
	doi = {10.1088/0264-9381/29/12/124013},
  
	year = 2012,
	month = {jun},
  
	publisher = {{IOP} Publishing},
  
	volume = {29},
  
	number = {12},
  
	pages = {124013},
  
	author = {B Sathyaprakash and M Abernathy and F Acernese and P Ajith and B Allen and P Amaro-Seoane and N Andersson and S Aoudia and K Arun and P Astone and B Krishnan and L Barack and F Barone and B Barr and M Barsuglia and M Bassan and R Bassiri and M Beker and N Beveridge and M Bizouard and C Bond and S Bose and L Bosi and S Braccini and C Bradaschia and M Britzger and F Brueckner and T Bulik and H J Bulten and O Burmeister and E Calloni and P Campsie and L Carbone and G Cella and E Chalkley and E Chassande-Mottin and S Chelkowski and A Chincarini and A Di Cintio and J Clark and E Coccia and C N Colacino and J Colas and A Colla and A Corsi and A Cumming and L Cunningham and E Cuoco and S Danilishin and K Danzmann and E Daw and R De Salvo and W Del Pozzo and T Dent and R De Rosa and L Di Fiore and M Di Paolo Emilio and A Di Virgilio and A Dietz and M Doets and J Dueck and M Edwards and V Fafone and S Fairhurst and P Falferi and M Favata and V Ferrari and F Ferrini and F Fidecaro and R Flaminio and J Franc and F Frasconi and A Freise and D Friedrich and P Fulda and J Gair and M Galimberti and G Gemme and E Genin and A Gennai and A Giazotto and K Glampedakis and S Gossan and R Gouaty and C Graef and W Graham and M Granata and H Grote and G Guidi and J Hallam and G Hammond and M Hannam and J Harms and K Haughian and I Hawke and D Heinert and M Hendry and I Heng and E Hennes and S Hild and J Hough and D Huet and S Husa and S Huttner and B Iyer and D I Jones and G Jones and I Kamaretsos and C Kant Mishra and F Kawazoe and F Khalili and B Kley and K Kokeyama and K Kokkotas and S Kroker and R Kumar and K Kuroda and B Lagrange and N Lastzka and T G F Li and M Lorenzini and G Losurdo and H Lück and E Majorana and V Malvezzi and I Mandel and V Mandic and S Marka and F Marin and F Marion and J Marque and I Martin and D Mc Leod and D Mckechan and M Mehmet and C Michel and Y Minenkov and N Morgado and A Morgia and S Mosca and L Moscatelli and B Mours and H Müller-Ebhardt and P Murray and L Naticchioni and R Nawrodt and J Nelson and R O' Shaughnessy and C D Ott and C Palomba and A Paoli and G Parguez and A Pasqualetti and R Passaquieti and D Passuello and M Perciballi and F Piergiovanni and L Pinard and M Pitkin and W Plastino and M Plissi and R Poggiani and P Popolizio and E Porter and M Prato and G Prodi and M Punturo and P Puppo and D Rabeling and I Racz and P Rapagnani and V Re and J Read and T Regimbau and H Rehbein and S Reid and F Ricci and F Richard and C Robinson and A Rocchi and R Romano and S Rowan and A Rüdiger and A Samblowski and L Santamar{\'{\i}
}a and B Sassolas and R Schilling and P Schmidt and R Schnabel and B Schutz and C Schwarz and J Scott and P Seidel and A M Sintes and K Somiya and C F Sopuerta and B Sorazu and F Speirits and L Storchi and K Strain and S Strigin and P Sutton and S Tarabrin and B Taylor and A Thürin and K Tokmakov and M Tonelli and H Tournefier and R Vaccarone and H Vahlbruch and J F J van den Brand and C Van Den Broeck and S van der Putten and M van Veggel and A Vecchio and J Veitch and F Vetrano and A Vicere and S Vyatchanin and P We{\ss}els and B Willke and W Winkler and G Woan and A Woodcraft and K Yamamoto},
  
	title = {Scientific objectives of Einstein Telescope},
  
	journal = {Classical and Quantum Gravity}
}

@article{Punturo_2010,
doi = {10.1088/0264-9381/27/8/084007},
url = {https://dx.doi.org/10.1088/0264-9381/27/8/084007},
year = {2010},
month = {apr},
publisher = {},
volume = {27},
number = {8},
pages = {084007},
author = {M Punturo and M Abernathy and F Acernese and B Allen and N Andersson and K Arun and F Barone and B Barr and M Barsuglia and M Beker and N Beveridge and S Birindelli and S Bose and L Bosi and S Braccini and C Bradaschia and T Bulik and E Calloni and G Cella and E Chassande Mottin and S Chelkowski and A Chincarini and J Clark and E Coccia and C Colacino and J Colas and A Cumming and L Cunningham and E Cuoco and S Danilishin and K Danzmann and G De Luca and R De Salvo and T Dent and R Derosa and L Di Fiore and A Di Virgilio and M Doets and V Fafone and P Falferi and R Flaminio and J Franc and F Frasconi and A Freise and P Fulda and J Gair and G Gemme and A Gennai and A Giazotto and K Glampedakis and M Granata and H Grote and G Guidi and G Hammond and M Hannam and J Harms and D Heinert and M Hendry and I Heng and E Hennes and S Hild and J Hough and S Husa and S Huttner and G Jones and F Khalili and K Kokeyama and K Kokkotas and B Krishnan and M Lorenzini and H Lück and E Majorana and I Mandel and V Mandic and I Martin and C Michel and Y Minenkov and N Morgado and S Mosca and B Mours and H Müller-Ebhardt and P Murray and R Nawrodt and J Nelson and R Oshaughnessy and C D Ott and C Palomba and A Paoli and G Parguez and A Pasqualetti and R Passaquieti and D Passuello and L Pinard and R Poggiani and P Popolizio and M Prato and P Puppo and D Rabeling and P Rapagnani and J Read and T Regimbau and H Rehbein and S Reid and L Rezzolla and F Ricci and F Richard and A Rocchi and S Rowan and A Rüdiger and B Sassolas and B Sathyaprakash and R Schnabel and C Schwarz and P Seidel and A Sintes and K Somiya and F Speirits and K Strain and S Strigin and P Sutton and S Tarabrin and J van den Brand and C van Leewen and M van Veggel and C van den Broeck and A Vecchio and J Veitch and F Vetrano and A Vicere and S Vyatchanin and B Willke and G Woan and P Wolfango and K Yamamoto},
title = {The third generation of gravitational wave observatories and their science reach},
journal = {Classical and Quantum Gravity},

}

@misc{Evans_2021,
      title={A Horizon Study for Cosmic Explorer: Science, Observatories, and Community}, 
      author={Matthew Evans and Rana X Adhikari and Chaitanya Afle and Stefan W. Ballmer and Sylvia Biscoveanu and Ssohrab Borhanian and Duncan A. Brown and Yanbei Chen and Robert Eisenstein and Alexandra Gruson and Anuradha Gupta and Evan D. Hall and Rachael Huxford and Brittany Kamai and Rahul Kashyap and Jeff S. Kissel and Kevin Kuns and Philippe Landry and Amber Lenon and Geoffrey Lovelace and Lee McCuller and Ken K. Y. Ng and Alexander H. Nitz and Jocelyn Read and B. S. Sathyaprakash and David H. Shoemaker and Bram J. J. Slagmolen and Joshua R. Smith and Varun Srivastava and Ling Sun and Salvatore Vitale and Rainer Weiss},
      year={2021},
      eprint={2109.09882},
      archivePrefix={arXiv},
      primaryClass={astro-ph.IM}
}

@article{Maartens_2021,
	doi = {10.1088/1475-7516/2021/12/009},
  
	year = 2021,
	month = {dec},
  
	publisher = {{IOP} Publishing},
  
	volume = {2021},
  
	number = {12},
  
	pages = {009},
  
	author = {Roy Maartens and Jos{\'{e}
} Fonseca and Stefano Camera and Sheean Jolicoeur and Jan-Albert Viljoen and Chris Clarkson},
  
	title = {Magnification and evolution biases in large-scale structure surveys},
  
	journal = {Journal of Cosmology and Astroparticle Physics}
}

@article{Finn_1996,
	doi = {10.1103/physrevd.53.2878},
  
	year = 1996,
	month = {mar},
  
	publisher = {American Physical Society ({APS})},
  
	volume = {53},
  
	number = {6},
  
	pages = {2878--2894},
  
	author = {Lee Samuel Finn},
  
	title = {Binary inspiral, gravitational radiation, and cosmology},
  
	journal = {Physical Review D}
}

@article{GWTC2,
	doi = {10.3847/2041-8213/abe949},

	year = 2021,
	month = {may},
  
	publisher = {American Astronomical Society},
  
	volume = {913},
  
	number = {1},
  
	pages = {L7},
  
	author = {R. Abbott and T. D. Abbott and S. Abraham and F. Acernese and K. Ackley and A. Adams and C. Adams and R. X. Adhikari and V. B. Adya and C. Affeldt and M. Agathos and K. Agatsuma and N. Aggarwal and O. D. Aguiar and L. Aiello and A. Ain and P. Ajith and G. Allen and A. Allocca and P. A. Altin and A. Amato and S. Anand and A. Ananyeva and S. B. Anderson and W. G. Anderson and S. V. Angelova and S. Ansoldi and J. M. Antelis and S. Antier and S. Appert and K. Arai and M. C. Araya and J. S. Areeda and M. Ar{\`{e}
}ne and N. Arnaud and S. M. Aronson and K. G. Arun and Y. Asali and S. Ascenzi and G. Ashton and S. M. Aston and P. Astone and F. Aubin and P. Aufmuth and K. AultONeal and C. Austin and V. Avendano and S. Babak and F. Badaracco and M. K. M. Bader and S. Bae and A. M. Baer and S. Bagnasco and J. Baird and M. Ball and G. Ballardin and S. W. Ballmer and A. Bals and A. Balsamo and G. Baltus and S. Banagiri and D. Bankar and R. S. Bankar and J. C. Barayoga and C. Barbieri and B. C. Barish and D. Barker and P. Barneo and S. Barnum and F. Barone and B. Barr and L. Barsotti and M. Barsuglia and D. Barta and J. Bartlett and I. Bartos and R. Bassiri and A. Basti and M. Bawaj and J. C. Bayley and M. Bazzan and B. R. Becher and B. B{\'{e}}csy and V. M. Bedakihale and M. Bejger and I. Belahcene and D. Beniwal and M. G. Benjamin and T. F. Bennett and J. D. Bentley and F. Bergamin and B. K. Berger and G. Bergmann and S. Bernuzzi and C. P. L. Berry and D. Bersanetti and A. Bertolini and J. Betzwieser and R. Bhandare and A. V. Bhandari and D. Bhattacharjee and J. Bidler and I. A. Bilenko and G. Billingsley and R. Birney and O. Birnholtz and S. Biscans and M. Bischi and S. Biscoveanu and A. Bisht and M. Bitossi and M.-A. Bizouard and J. K. Blackburn and J. Blackman and C. D. Blair and D. G. Blair and R. M. Blair and O. Blanch and F. Bobba and N. Bode and M. Boer and Y. Boetzel and G. Bogaert and M. Boldrini and F. Bondu and E. Bonilla and R. Bonnand and P. Booker and B. A. Boom and R. Bork and V. Boschi and S. Bose and V. Bossilkov and V. Boudart and Y. Bouffanais and A. Bozzi and C. Bradaschia and P. R. Brady and A. Bramley and M. Branchesi and J. E. Brau and M. Breschi and T. Briant and J. H. Briggs and F. Brighenti and A. Brillet and M. Brinkmann and P. Brockill and A. F. Brooks and J. Brooks and D. D. Brown and S. Brunett and G. Bruno and R. Bruntz and A. Buikema and T. Bulik and H. J. Bulten and A. Buonanno and R. Buscicchio and D. Buskulic and R. L. Byer and M. Cabero and L. Cadonati and M. Caesar and G. Cagnoli and C. Cahillane and J. Calder{\'{o}}n Bustillo and J. D. Callaghan and T. A. Callister and E. Calloni and J. B. Camp and M. Canepa and K. C. Cannon and H. Cao and J. Cao and G. Carapella and F. Carbognani and M. F. Carney and M. Carpinelli and G. Carullo and T. L. Carver and J. Casanueva Diaz and C. Casentini and S. Caudill and M. Cavagli{\`{a}} and F. Cavalier and R. Cavalieri and G. Cella and P. Cerd{\'{a}}-Dur{\'{a}}n and E. Cesarini and W. Chaibi and K. Chakravarti and C.-L. Chan and C. Chan and K. Chandra and P. Chanial and S. Chao and P. Charlton and E. A. Chase and E. Chassande-Mottin and D. Chatterjee and D. Chattopadhyay and M. Chaturvedi and K. Chatziioannou and A. Chen and H. Y. Chen and X. Chen and Y. Chen and H.-P. Cheng and C. K. Cheong and H. Y. Chia and F. Chiadini and R. Chierici and A. Chincarini and A. Chiummo and G. Cho and H. S. Cho and M. Cho and S. Choate and N. Christensen and Q. Chu and S. Chua and K. W. Chung and S. Chung and G. Ciani and P. Ciecielag and M. Cie{\'{s}}lar and M. Cifaldi and A. A. Ciobanu and R. Ciolfi and F. Cipriano and A. Cirone and F. Clara and E. N. Clark and J. A. Clark and L. Clarke and P. Clearwater and S. Clesse and F. Cleva and E. Coccia and P.-F. Cohadon and D. E. Cohen and M. Colleoni and C. G. Collette and C. Collins and M. Colpi and M. Constancio and L. Conti and S. J. Cooper and P. Corban and T. R. Corbitt and I. Cordero-Carri{\'{o}}n and S. Corezzi and K. R. Corley and N. Cornish and D. Corre and A. Corsi and S. Cortese and C. A. Costa and R. Cotesta and M. W. Coughlin and S. B. Coughlin and J.-P. Coulon and S. T. Countryman and P. Couvares and P. B. Covas and D. M. Coward and M. J. Cowart and D. C. Coyne and R. Coyne and J. D. E. Creighton and T. D. Creighton and M. Croquette and S. G. Crowder and J. R. Cudell and T. J. Cullen and A. Cumming and R. Cummings and L. Cunningham and E. Cuoco and M. Curylo and T. Dal Canton and G. D{\'{a}}lya and A. Dana and L. M. DaneshgaranBajastani and B. D'Angelo and S. L. Danilishin and S. D'Antonio and K. Danzmann and C. Darsow-Fromm and A. Dasgupta and L. E. H. Datrier and V. Dattilo and I. Dave and M. Davier and G. S. Davies and D. Davis and E. J. Daw and R. Dean and D. DeBra and M. Deenadayalan and J. Degallaix and M. De Laurentis and S. Del{\'{e}}glise and V. Del Favero and F. De Lillo and N. De Lillo and W. Del Pozzo and L. M. DeMarchi and F. De Matteis and V. D'Emilio and N. Demos and T. Denker and T. Dent and A. Depasse and R. De Pietri and R. De Rosa and C. De Rossi and R. DeSalvo and O. de Varona and S. Dhurandhar and M. C. D{\'{\i}}az and M. Diaz-Ortiz and N. A. Didio and T. Dietrich and L. Di Fiore and C. DiFronzo and C. Di Giorgio and F. Di Giovanni and M. Di Giovanni and T. Di Girolamo and A. Di Lieto and B. Ding and S. Di Pace and I. Di Palma and F. Di Renzo and A. K. Divakarla and A. Dmitriev and Z. Doctor and L. D'Onofrio and F. Donovan and K. L. Dooley and S. Doravari and I. Dorrington and T. P. Downes and M. Drago and J. C. Driggers and Z. Du and J.-G. Ducoin and P. Dupej and O. Durante and D. D'Urso and P.-A. Duverne and S. E. Dwyer and P. J. Easter and G. Eddolls and B. Edelman and T. B. Edo and O. Edy and A. Effler and J. Eichholz and S. S. Eikenberry and M. Eisenmann and R. A. Eisenstein and A. Ejlli and L. Errico and R. C. Essick and H. Estell{\'{e}}s and D. Estevez and Z. B. Etienne and T. Etzel and M. Evans and T. M. Evans and B. E. Ewing and V. Fafone and H. Fair and S. Fairhurst and X. Fan and A. M. Farah and S. Farinon and B. Farr and W. M. Farr and E. J. Fauchon-Jones and M. Favata and M. Fays and M. Fazio and J. Feicht and M. M. Fejer and F. Feng and E. Fenyvesi and D. L. Ferguson and A. Fernandez-Galiana and I. Ferrante and T. A. Ferreira and F. Fidecaro and P. Figura and I. Fiori and D. Fiorucci and M. Fishbach and R. P. Fisher and J. M. Fishner and R. Fittipaldi and M. Fitz-Axen and V. Fiumara and R. Flaminio and E. Floden and E. Flynn and H. Fong and J. A. Font and P. W. F. Forsyth and J.-D. Fournier and S. Frasca and F. Frasconi and Z. Frei and A. Freise and R. Frey and V. Frey and P. Fritschel and V. V. Frolov and G. G. Fronz{\'{e}} and P. Fulda and M. Fyffe and H. A. Gabbard and B. U. Gadre and S. M. Gaebel and J. R. Gair and J. Gais and S. Galaudage and R. Gamba and D. Ganapathy and A. Ganguly and S. G. Gaonkar and B. Garaventa and C. Garc{\'{\i}}a-Quir{\'{o}}s and F. Garufi and B. Gateley and S. Gaudio and V. Gayathri and G. Gemme and A. Gennai and D. George and J. George and L. Gergely and S. Ghonge and Abhirup Ghosh and Archisman Ghosh and S. Ghosh and B. Giacomazzo and L. Giacoppo and J. A. Giaime and K. D. Giardina and D. R. Gibson and C. Gier and K. Gill and P. Giri and J. Glanzer and A. E. Gleckl and P. Godwin and E. Goetz and R. Goetz and N. Gohlke and B. Goncharov and G. Gonz{\'{a}}lez and A. Gopakumar and S. E. Gossan and M. Gosselin and R. Gouaty and B. Grace and A. Grado and M. Granata and V. Granata and A. Grant and S. Gras and P. Grassia and C. Gray and R. Gray and G. Greco and A. C. Green and R. Green and E. M. Gretarsson and H. L. Griggs and G. Grignani and A. Grimaldi and E. Grimes and S. J. Grimm and H. Grote and S. Grunewald and P. Gruning and J. G. Guerrero and G. M. Guidi and A. R. Guimaraes and G. Guix{\'{e}} and H. K. Gulati and Y. Guo and Anchal Gupta and Anuradha Gupta and P. Gupta and E. K. Gustafson and R. Gustafson and F. Guzman and L. Haegel and O. Halim and E. D. Hall and E. Z. Hamilton and G. Hammond and M. Haney and M. M. Hanke and J. Hanks and C. Hanna and O. A. Hannuksela and O. Hannuksela and H. Hansen and T. J. Hansen and J. Hanson and T. Harder and T. Hardwick and K. Haris and J. Harms and G. M. Harry and I. W. Harry and D. Hartwig and R. K. Hasskew and C.-J. Haster and K. Haughian and F. J. Hayes and J. Healy and A. Heidmann and M. C. Heintze and J. Heinze and J. Heinzel and H. Heitmann and F. Hellman and P. Hello and A. F. Helmling-Cornell and G. Hemming and M. Hendry and I. S. Heng and E. Hennes and J. Hennig and M. H. Hennig and F. Hernandez Vivanco and M. Heurs and S. Hild and P. Hill and A. S. Hines and S. Hochheim and E. Hofgard and D. Hofman and J. N. Hohmann and A. M. Holgado and N. A. Holland and I. J. Hollows and Z. J. Holmes and K. Holt and D. E. Holz and P. Hopkins and C. Horst and J. Hough and E. J. Howell and C. G. Hoy and D. Hoyland and Y. Huang and M. T. Hübner and A. D. Huddart and E. A. Huerta and B. Hughey and V. Hui and S. Husa and S. H. Huttner and B. M. Hutzler and R. Huxford and T. Huynh-Dinh and B. Idzkowski and A. Iess and S. Imperato and H. Inchauspe and C. Ingram and G. Intini and M. Isi and B. R. Iyer and V. JaberianHamedan and T. Jacqmin and S. J. Jadhav and S. P. Jadhav and A. L. James and K. Jani and K. Janssens and N. N. Janthalur and P. Jaranowski and D. Jariwala and R. Jaume and A. C. Jenkins and M. Jeunon and J. Jiang and G. R. Johns and A. W. Jones and D. I. Jones and J. D. Jones and P. Jones and R. Jones and R. J. G. Jonker and L. Ju and J. Junker and C. V. Kalaghatgi and V. Kalogera and B. Kamai and S. Kandhasamy and G. Kang and J. B. Kanner and S. J. Kapadia and D. P. Kapasi and C. Karathanasis and S. Karki and R. Kashyap and M. Kasprzack and W. Kastaun and S. Katsanevas and E. Katsavounidis and W. Katzman and K. Kawabe and F. K{\'{e}}f{\'{e}}lian and D. Keitel and J. S. Key and S. Khadka and F. Y. Khalili and I. Khan and S. Khan and E. A. Khazanov and N. Khetan and M. Khursheed and N. Kijbunchoo and C. Kim and G. J. Kim and J. C. Kim and K. Kim and W. S. Kim and Y.-M. Kim and C. Kimball and P. J. King and M. Kinley-Hanlon and R. Kirchhoff and J. S. Kissel and L. Kleybolte and S. Klimenko and T. D. Knowles and E. Knyazev and P. Koch and S. M. Koehlenbeck and G. Koekoek and S. Koley and M. Kolstein and K. Komori and V. Kondrashov and A. Kontos and N. Koper and M. Korobko and W. Z. Korth and M. Kovalam and D. B. Kozak and C. Krämer and V. Kringel and N. V. Krishnendu and A. Kr{\'{o}}lak and G. Kuehn and A. Kumar and P. Kumar and Rahul Kumar and Rakesh Kumar and K. Kuns and S. Kwang and B. D. Lackey and D. Laghi and E. Lalande and T. L. Lam and A. Lamberts and M. Landry and B. B. Lane and R. N. Lang and J. Lange and B. Lantz and R. K. Lanza and I. La Rosa and A. Lartaux-Vollard and P. D. Lasky and M. Laxen and A. Lazzarini and C. Lazzaro and P. Leaci and S. Leavey and Y. K. Lecoeuche and H. M. Lee and H. W. Lee and J. Lee and K. Lee and J. Lehmann and E. Leon and N. Leroy and N. Letendre and Y. Levin and A. Li and J. Li and K. J. L. Li and T. G. F. Li and X. Li and F. Linde and S. D. Linker and J. N. Linley and T. B. Littenberg and J. Liu and X. Liu and M. Llorens-Monteagudo and R. K. L. Lo and A. Lockwood and L. T. London and A. Longo and M. Lorenzini and V. Loriette and M. Lormand and G. Losurdo and J. D. Lough and C. O. Lousto and G. Lovelace and H. Lück and D. Lumaca and A. P. Lundgren and Y. Ma and R. Macas and M. MacInnis and D. M. Macleod and I. A. O. MacMillan and A. Macquet and I. Maga{\~{n}}a Hernandez and F. Maga{\~{n}}a-Sandoval and C. Magazz{\`{u}} and R. M. Magee and E. Majorana and I. Maksimovic and S. Maliakal and A. Malik and N. Man and V. Mandic and V. Mangano and G. L. Mansell and M. Manske and M. Mantovani and M. Mapelli and F. Marchesoni and F. Marion and S. M{\'{a}}rka and Z. M{\'{a}}rka and C. Markakis and A. S. Markosyan and A. Markowitz and E. Maros and A. Marquina and S. Marsat and F. Martelli and I. W. Martin and R. M. Martin and M. Martinez and V. Martinez and D. V. Martynov and H. Masalehdan and K. Mason and E. Massera and A. Masserot and T. J. Massinger and M. Masso-Reid and S. Mastrogiovanni and A. Matas and M. Mateu-Lucena and F. Matichard and M. Matiushechkina and N. Mavalvala and E. Maynard and J. J. McCann and R. McCarthy and D. E. McClelland and S. McCormick and L. McCuller and S. C. McGuire and C. McIsaac and J. McIver and D. J. McManus and T. McRae and S. T. McWilliams and D. Meacher and G. D. Meadors and M. Mehmet and A. K. Mehta and A. Melatos and D. A. Melchor and G. Mendell and A. Menendez-Vazquez and R. A. Mercer and L. Mereni and K. Merfeld and E. L. Merilh and J. D. Merritt and M. Merzougui and S. Meshkov and C. Messenger and C. Messick and R. Metzdorff and P. M. Meyers and F. Meylahn and A. Mhaske and A. Miani and H. Miao and I. Michaloliakos and C. Michel and H. Middleton and L. Milano and A. L. Miller and S. Miller and M. Millhouse and J. C. Mills and E. Milotti and M. C. Milovich-Goff and O. Minazzoli and Y. Minenkov and Ll. M. Mir and A. Mishkin and C. Mishra and T. Mistry and S. Mitra and V. P. Mitrofanov and G. Mitselmakher and R. Mittleman and G. Mo and K. Mogushi and S. R. P. Mohapatra and S. R. Mohite and I. Molina and M. Molina-Ruiz and M. Mondin and M. Montani and C. J. Moore and D. Moraru and F. Morawski and G. Moreno and S. Morisaki and B. Mours and C. M. Mow-Lowry and S. Mozzon and F. Muciaccia and Arunava Mukherjee and D. Mukherjee and Soma Mukherjee and Subroto Mukherjee and N. Mukund and A. Mullavey and J. Munch and E. A. Mu{\~{n}}iz and P. G. Murray and S. L. Nadji and A. Nagar and I. Nardecchia and L. Naticchioni and R. K. Nayak and B. F. Neil and J. Neilson and G. Nelemans and T. J. N. Nelson and M. Nery and A. Neunzert and K. Y. Ng and S. Ng and C. Nguyen and P. Nguyen and T. Nguyen and S. A. Nichols and S. Nissanke and F. Nocera and M. Noh and C. North and D. Nothard and L. K. Nuttall and J. Oberling and B. D. O'Brien and J. O'Dell and G. Oganesyan and G. H. Ogin and J. J. Oh and S. H. Oh and F. Ohme and H. Ohta and M. A. Okada and C. Olivetto and P. Oppermann and R. J. Oram and B. O'Reilly and R. G. Ormiston and N. Ormsby and L. F. Ortega and R. O'Shaughnessy and S. Ossokine and C. Osthelder and D. J. Ottaway and H. Overmier and B. J. Owen and A. E. Pace and G. Pagano and M. A. Page and G. Pagliaroli and A. Pai and S. A. Pai and J. R. Palamos and O. Palashov and C. Palomba and H. Pan and P. K. Panda and T. H. Pang and C. Pankow and F. Pannarale and B. C. Pant and F. Paoletti and A. Paoli and A. Paolone and W. Parker and D. Pascucci and A. Pasqualetti and R. Passaquieti and D. Passuello and M. Patel and B. Patricelli and E. Payne and T. C. Pechsiri and M. Pedraza and M. Pegoraro and A. Pele and S. Penn and A. Perego and C. J. Perez and C. P{\'{e}}rigois and A. Perreca and S. Perri{\`{e}}s and J. Petermann and D. Petterson and H. P. Pfeiffer and K. A. Pham and K. S. Phukon and O. J. Piccinni and M. Pichot and M. Piendibene and F. Piergiovanni and L. Pierini and V. Pierro and G. Pillant and F. Pilo and L. Pinard and I. M. Pinto and K. Piotrzkowski and M. Pirello and M. Pitkin and E. Placidi and W. Plastino and C. Pluchar and R. Poggiani and E. Polini and D. Y. T. Pong and S. Ponrathnam and P. Popolizio and E. K. Porter and A. Poverman and J. Powell and M. Pracchia and A. K. Prajapati and K. Prasai and R. Prasanna and G. Pratten and T. Prestegard and M. Principe and G. A. Prodi and L. Prokhorov and P. Prosposito and A. Puecher and M. Punturo and F. Puosi and P. Puppo and M. Pürrer and H. Qi and V. Quetschke and P. J. Quinonez and R. Quitzow-James and F. J. Raab and G. Raaijmakers and H. Radkins and N. Radulesco and P. Raffai and H. Rafferty and S. X. Rail and S. Raja and C. Rajan and B. Rajbhandari and M. Rakhmanov and K. E. Ramirez and T. D. Ramirez and A. Ramos-Buades and J. Rana and K. Rao and P. Rapagnani and U. D. Rapol and B. Ratto and V. Raymond and M. Razzano and J. Read and T. Regimbau and L. Rei and S. Reid and D. H. Reitze and P. Rettegno and F. Ricci and C. J. Richardson and J. W. Richardson and L. Richardson and P. M. Ricker and G. Riemenschneider and K. Riles and M. Rizzo and N. A. Robertson and F. Robinet and A. Rocchi and J. A. Rocha and S. Rodriguez and R. D. Rodriguez-Soto and L. Rolland and J. G. Rollins and V. J. Roma and M. Romanelli and R. Romano and C. L. Romel and A. Romero and I. M. Romero-Shaw and J. H. Romie and S. Ronchini and C. A. Rose and D. Rose and K. Rose and M. J. B. Rosell and D. Rosi{\'{n}}ska and S. G. Rosofsky and M. P. Ross and S. Rowan and S. J. Rowlinson and Santosh Roy and Soumen Roy and P. Ruggi and K. Ryan and S. Sachdev and T. Sadecki and M. Sakellariadou and O. S. Salafia and L. Salconi and M. Saleem and A. Samajdar and E. J. Sanchez and J. H. Sanchez and L. E. Sanchez and N. Sanchis-Gual and J. R. Sanders and K. A. Santiago and E. Santos and T. R. Saravanan and N. Sarin and B. Sassolas and B. S. Sathyaprakash and O. Sauter and R. L. Savage and V. Savant and D. Sawant and S. Sayah and D. Schaetzl and P. Schale and M. Scheel and J. Scheuer and A. Schindler-Tyka and P. Schmidt and R. Schnabel and R. M. S. Schofield and A. Schönbeck and E. Schreiber and B. W. Schulte and B. F. Schutz and O. Schwarm and E. Schwartz and J. Scott and S. M. Scott and M. Seglar-Arroyo and E. Seidel and D. Sellers and A. S. Sengupta and N. Sennett and D. Sentenac and V. Sequino and A. Sergeev and Y. Setyawati and T. Shaffer and M. S. Shahriar and S. Sharifi and A. Sharma and P. Sharma and P. Shawhan and H. Shen and M. Shikauchi and R. Shink and D. H. Shoemaker and D. M. Shoemaker and K. Shukla and S. ShyamSundar and M. Sieniawska and D. Sigg and L. P. Singer and D. Singh and N. Singh and A. Singha and A. Singhal and A. M. Sintes and V. Sipala and V. Skliris and B. J. J. Slagmolen and T. J. Slaven-Blair and J. Smetana and J. R. Smith and R. J. E. Smith and S. N. Somala and E. J. Son and S. Soni and B. Sorazu and V. Sordini and F. Sorrentino and N. Sorrentino and R. Soulard and T. Souradeep and E. Sowell and A. P. Spencer and M. Spera and A. K. Srivastava and V. Srivastava and K. Staats and C. Stachie and D. A. Steer and M. Steinke and J. Steinlechner and S. Steinlechner and D. Steinmeyer and S. P. Stevenson and G. Stolle-McAllister and D. J. Stops and M. Stover and K. A. Strain and G. Stratta and A. Strunk and R. Sturani and A. L. Stuver and J. Südbeck and S. Sudhagar and V. Sudhir and H. G. Suh and T. Z. Summerscales and H. Sun and L. Sun and S. Sunil and A. Sur and J. Suresh and P. J. Sutton and B. L. Swinkels and M. J. Szczepa{\'{n}}czyk and M. Tacca and S. C. Tait and C. Talbot and A. J. Tanasijczuk and D. B. Tanner and D. Tao and A. Tapia and E. N. Tapia San Martin and J. D. Tasson and R. Taylor and R. Tenorio and L. Terkowski and M. P. Thirugnanasambandam and L. Thomas and M. Thomas and P. Thomas and J. E. Thompson and S. R. Thondapu and K. A. Thorne and E. Thrane and Shubhanshu Tiwari and Srishti Tiwari and V. Tiwari and K. Toland and A. E. Tolley and M. Tonelli and Z. Tornasi and A. Torres-Forn{\'{e}} and C. I. Torrie and I. Tosta e Melo and D. Töyrä and A. T. Tran and A. Trapananti and F. Travasso and G. Traylor and M. C. Tringali and A. Tripathee and A. Trovato and R. J. Trudeau and D. S. Tsai and K. W. Tsang and M. Tse and R. Tso and L. Tsukada and D. Tsuna and T. Tsutsui and M. Turconi and A. S. Ubhi and R. P. Udall and K. Ueno and D. Ugolini and C. S. Unnikrishnan and A. L. Urban and S. A. Usman and A. C. Utina and H. Vahlbruch and G. Vajente and A. Vajpeyi and G. Valdes and M. Valentini and V. Valsan and N. van Bakel and M. van Beuzekom and J. F. J. van den Brand and C. Van Den Broeck and D. C. Vander-Hyde and L. van der Schaaf and J. V. van Heijningen and M. Vardaro and A. F. Vargas and V. Varma and S. Vass and M. Vas{\'{u}}th and A. Vecchio and G. Vedovato and J. Veitch and P. J. Veitch and K. Venkateswara and J. Venneberg and G. Venugopalan and D. Verkindt and Y. Verma and D. Veske and F. Vetrano and A. Vicer{\'{e}} and A. D. Viets and V. Villa-Ortega and J.-Y. Vinet and S. Vitale and T. Vo and H. Vocca and C. Vorvick and S. P. Vyatchanin and A. R. Wade and L. E. Wade and M. Wade and R. C. Walet and M. Walker and G. S. Wallace and L. Wallace and S. Walsh and J. Z. Wang and S. Wang and W. H. Wang and Y. F. Wang and R. L. Ward and J. Warner and M. Was and N. Y. Washington and J. Watchi and B. Weaver and L. Wei and M. Weinert and A. J. Weinstein and R. Weiss and F. Wellmann and L. Wen and P. We{\ss}els and J. W. Westhouse and K. Wette and J. T. Whelan and D. D. White and L. V. White and B. F. Whiting and C. Whittle and D. M. Wilken and D. Williams and M. J. Williams and A. R. Williamson and J. L. Willis and B. Willke and D. J. Wilson and M. H. Wimmer and W. Winkler and C. C. Wipf and G. Woan and J. Woehler and J. K. Wofford and I. C. F. Wong and J. Wrangel and J. L. Wright and D. S. Wu and D. M. Wysocki and L. Xiao and H. Yamamoto and L. Yang and Y. Yang and Z. Yang and M. J. Yap and D. W. Yeeles and A. Yoon and Hang Yu and Haocun Yu and S. H. R. Yuen and A. Zadro{\.{z}}ny and M. Zanolin and T. Zelenova and J.-P. Zendri and M. Zevin and J. Zhang and L. Zhang and R. Zhang and T. Zhang and C. Zhao and G. Zhao and M. Zhou and Z. Zhou and X. J. Zhu and A. B. Zimmerman and M. E. Zucker and J. Zweizig},
  
	title = {Population Properties of Compact Objects from the Second {LIGO}{\textendash}Virgo Gravitational-Wave Transient Catalog},
  
	journal = {The Astrophysical Journal Letters}
}

@article{Namikawa,
	doi = {10.1088/1475-7516/2021/01/036},
  
	year = 2021,
	month = {jan},
  
	publisher = {{IOP} Publishing},
  
	volume = {2021},
  
	number = {01},
  
	pages = {036--036},
  
	author = {Toshiya Namikawa},
  
	title = {Analyzing clustering of astrophysical gravitational-wave sources: luminosity-distance space distortions},
  
	journal = {Journal of Cosmology and Astroparticle Physics}
}

@article{Zazzera_2024,
   title={Magnification and evolution bias of transient sources: GWs and SNIa},
   volume={2024},
   ISSN={1475-7516},
   url={http://dx.doi.org/10.1088/1475-7516/2024/05/095},
   DOI={10.1088/1475-7516/2024/05/095},
   number={05},
   journal={Journal of Cosmology and Astroparticle Physics},
   publisher={IOP Publishing},
   author={Zazzera, Stefano and Fonseca, José and Baker, Tessa and Clarkson, Chris},
   year={2024},
   month=may, pages={095} }

@misc{LSST_2012,
      title={Large Synoptic Survey Telescope: Dark Energy Science Collaboration}, 
      author={{LSST Dark Energy Science Collaboration}},
      year={2012},
      eprint={1211.0310},
      archivePrefix={arXiv},
      primaryClass={astro-ph.CO},
      url={https://arxiv.org/abs/1211.0310}, 
}

@article{Ma_2006,
   title={Effects of Photometric Redshift Uncertainties on Weak‐Lensing Tomography},
   volume={636},
   ISSN={1538-4357},
   url={http://dx.doi.org/10.1086/497068},
   DOI={10.1086/497068},
   number={1},
   journal={The Astrophysical Journal},
   publisher={American Astronomical Society},
   author={Ma, Zhaoming and Hu, Wayne and Huterer, Dragan},
   year={2006},
   month=jan, pages={21–29} }

@article{Viljoen_2021,
   title={Multi-wavelength spectroscopic probes: prospects for primordial non-Gaussianity and relativistic effects},
   volume={2021},
   ISSN={1475-7516},
   url={http://dx.doi.org/10.1088/1475-7516/2021/11/010},
   DOI={10.1088/1475-7516/2021/11/010},
   number={11},
   journal={Journal of Cosmology and Astroparticle Physics},
   publisher={IOP Publishing},
   author={Viljoen, Jan-Albert and Fonseca, José and Maartens, Roy},
   year={2021},
   month=nov, pages={010} }

@article{Abramo_2022,
   title={Fisher matrix for the angular power spectrum of multi-tracer galaxy surveys},
   volume={2022},
   ISSN={1475-7516},
   url={http://dx.doi.org/10.1088/1475-7516/2022/08/073},
   DOI={10.1088/1475-7516/2022/08/073},
   number={08},
   journal={Journal of Cosmology and Astroparticle Physics},
   publisher={IOP Publishing},
   author={Abramo, L. Raul and Dinarte Ferri, João Vitor and Tashiro, Ian Lucas and Loureiro, Arthur},
   year={2022},
   month=aug, pages={073} }

@article{Mastrogiovanni_24,
author = {Mastrogiovanni, Simone and Karathanasis, Christos and Gair, Jonathan and Ashton, Gregory and Rinaldi, Stefano and Huang, Hsiang-Yu and Dálya, Gergely},
title = {Cosmology with Gravitational Waves: A Review},
journal = {Annalen der Physik},
volume = {536},
number = {2},
pages = {2200180},
doi = {https://doi.org/10.1002/andp.202200180},
url = {https://onlinelibrary.wiley.com/doi/abs/10.1002/andp.202200180},
eprint = {https://onlinelibrary.wiley.com/doi/pdf/10.1002/andp.202200180},
year = {2024}
}

@article{Planck_2018,
   title={Planck2018 results: VI. Cosmological parameters},
   volume={641},
   ISSN={1432-0746},
   url={http://dx.doi.org/10.1051/0004-6361/201833910},
   DOI={10.1051/0004-6361/201833910},
   journal={Astronomy &amp; Astrophysics},
   publisher={EDP Sciences},
   author={{Planck Collaboration}, Aghanim, N. and Akrami, Y. and Ashdown, M. and Aumont, J. and Baccigalupi, C. and Ballardini, M. and Banday, A. J. and Barreiro, R. B. and Bartolo, N. and Basak, S. and Battye, R. and Benabed, K. and Bernard, J.-P. and Bersanelli, M. and Bielewicz, P. and Bock, J. J. and Bond, J. R. and Borrill, J. and Bouchet, F. R. and Boulanger, F. and Bucher, M. and Burigana, C. and Butler, R. C. and Calabrese, E. and Cardoso, J.-F. and Carron, J. and Challinor, A. and Chiang, H. C. and Chluba, J. and Colombo, L. P. L. and Combet, C. and Contreras, D. and Crill, B. P. and Cuttaia, F. and de Bernardis, P. and de Zotti, G. and Delabrouille, J. and Delouis, J.-M. and Di Valentino, E. and Diego, J. M. and Doré, O. and Douspis, M. and Ducout, A. and Dupac, X. and Dusini, S. and Efstathiou, G. and Elsner, F. and Enßlin, T. A. and Eriksen, H. K. and Fantaye, Y. and Farhang, M. and Fergusson, J. and Fernandez-Cobos, R. and Finelli, F. and Forastieri, F. and Frailis, M. and Fraisse, A. A. and Franceschi, E. and Frolov, A. and Galeotta, S. and Galli, S. and Ganga, K. and Génova-Santos, R. T. and Gerbino, M. and Ghosh, T. and González-Nuevo, J. and Górski, K. M. and Gratton, S. and Gruppuso, A. and Gudmundsson, J. E. and Hamann, J. and Handley, W. and Hansen, F. K. and Herranz, D. and Hildebrandt, S. R. and Hivon, E. and Huang, Z. and Jaffe, A. H. and Jones, W. C. and Karakci, A. and Keihänen, E. and Keskitalo, R. and Kiiveri, K. and Kim, J. and Kisner, T. S. and Knox, L. and Krachmalnicoff, N. and Kunz, M. and Kurki-Suonio, H. and Lagache, G. and Lamarre, J.-M. and Lasenby, A. and Lattanzi, M. and Lawrence, C. R. and Le Jeune, M. and Lemos, P. and Lesgourgues, J. and Levrier, F. and Lewis, A. and Liguori, M. and Lilje, P. B. and Lilley, M. and Lindholm, V. and López-Caniego, M. and Lubin, P. M. and Ma, Y.-Z. and Macías-Pérez, J. F. and Maggio, G. and Maino, D. and Mandolesi, N. and Mangilli, A. and Marcos-Caballero, A. and Maris, M. and Martin, P. G. and Martinelli, M. and Martínez-González, E. and Matarrese, S. and Mauri, N. and McEwen, J. D. and Meinhold, P. R. and Melchiorri, A. and Mennella, A. and Migliaccio, M. and Millea, M. and Mitra, S. and Miville-Deschênes, M.-A. and Molinari, D. and Montier, L. and Morgante, G. and Moss, A. and Natoli, P. and Nørgaard-Nielsen, H. U. and Pagano, L. and Paoletti, D. and Partridge, B. and Patanchon, G. and Peiris, H. V. and Perrotta, F. and Pettorino, V. and Piacentini, F. and Polastri, L. and Polenta, G. and Puget, J.-L. and Rachen, J. P. and Reinecke, M. and Remazeilles, M. and Renzi, A. and Rocha, G. and Rosset, C. and Roudier, G. and Rubiño-Martín, J. A. and Ruiz-Granados, B. and Salvati, L. and Sandri, M. and Savelainen, M. and Scott, D. and Shellard, E. P. S. and Sirignano, C. and Sirri, G. and Spencer, L. D. and Sunyaev, R. and Suur-Uski, A.-S. and Tauber, J. A. and Tavagnacco, D. and Tenti, M. and Toffolatti, L. and Tomasi, M. and Trombetti, T. and Valenziano, L. and Valiviita, J. and Van Tent, B. and Vibert, L. and Vielva, P. and Villa, F. and Vittorio, N. and Wandelt, B. D. and Wehus, I. K. and White, M. and White, S. D. M. and Zacchei, A. and Zonca, A.},
   year={2020},
   month=sep, pages={A6} }

@article{Afroz:2024joi,
    author = "Afroz, Samsuzzaman and Mukherjee, Suvodip",
    title = "{Prospect of precision cosmology and testing general relativity using binary black holes \textendash{} galaxies cross-correlation}",
    eprint = "2407.09262",
    archivePrefix = "arXiv",
    primaryClass = "astro-ph.CO",
    doi = "10.1093/mnras/stae2139",
    journal = "Mon. Not. Roy. Astron. Soc.",
    volume = "534",
    number = "2",
    pages = "1283--1298",
    year = "2024"
}

@article{Mukherjee:2022afz,
    author = "Mukherjee, Suvodip and Krolewski, Alex and Wandelt, Benjamin D. and Silk, Joseph",
    title = "{Cross-correlating dark sirens and galaxies: constraints on $H_0$ from GWTC-3 of LIGO-Virgo-KAGRA}",
    eprint = "2203.03643",
    archivePrefix = "arXiv",
    primaryClass = "astro-ph.CO",
    doi = "10.3847/1538-4357/ad7d90",
    journal = "Astrophys. J.",
    volume = "975",
    number = "2",
    pages = "189",
    year = "2024"
}

@article{Mukherjee:2019wfw,
    author = "Mukherjee, Suvodip and Wandelt, Benjamin D. and Silk, Joseph",
    title = "{Multimessenger tests of gravity with weakly lensed gravitational waves}",
    eprint = "1908.08950",
    archivePrefix = "arXiv",
    primaryClass = "astro-ph.CO",
    doi = "10.1103/PhysRevD.101.103509",
    journal = "Phys. Rev. D",
    volume = "101",
    number = "10",
    pages = "103509",
    year = "2020"
}

@article{Mukherjee:2019wcg,
    author = "Mukherjee, Suvodip and Wandelt, Benjamin D. and Silk, Joseph",
    title = "{Probing the theory of gravity with gravitational lensing of gravitational waves and galaxy surveys}",
    eprint = "1908.08951",
    archivePrefix = "arXiv",
    primaryClass = "astro-ph.CO",
    doi = "10.1093/mnras/staa827",
    journal = "Mon. Not. Roy. Astron. Soc.",
    volume = "494",
    number = "2",
    pages = "1956--1970",
    year = "2020"
}

@misc{mukherjee2018,
      title={Beyond the classical distance-redshift test: cross-correlating redshift-free standard candles and sirens with redshift surveys}, 
      author={Suvodip Mukherjee and Benjamin D. Wandelt},
      year={2018},
      eprint={1808.06615},
      archivePrefix={arXiv},
      primaryClass={astro-ph.CO},
      url={https://arxiv.org/abs/1808.06615}, 
}

@article{KAGRA:2013rdx,
    author = "Abbott, B. P. and others",
    collaboration = "KAGRA, LIGO Scientific, Virgo",
    title = "{Prospects for observing and localizing gravitational-wave transients with Advanced LIGO, Advanced Virgo and KAGRA}",
    eprint = "1304.0670",
    archivePrefix = "arXiv",
    primaryClass = "gr-qc",
    reportNumber = "LIGO-P1200087, VIR-0288A-12, JGW-P1808427",
    doi = "10.1007/s41114-020-00026-9",
    journal = "Living Rev. Rel.",
    volume = "19",
    pages = "1",
    year = "2016"
}

@article{AdvVirgo,
   title={Advanced Virgo: a second-generation interferometric gravitational wave detector},
   volume={32},
   ISSN={1361-6382},
   url={http://dx.doi.org/10.1088/0264-9381/32/2/024001},
   DOI={10.1088/0264-9381/32/2/024001},
   number={2},
   journal={Classical and Quantum Gravity},
   publisher={IOP Publishing},
   author={Acernese, F and Agathos, M and Agatsuma, K and Aisa, D and Allemandou, N and Allocca, A and Amarni, J and Astone, P and Balestri, G and Ballardin, G and Barone, F and Baronick, J-P and Barsuglia, M and Basti, A and Basti, F and Bauer, Th S and Bavigadda, V and Bejger, M and Beker, M G and Belczynski, C and Bersanetti, D and Bertolini, A and Bitossi, M and Bizouard, M A and Bloemen, S and Blom, M and Boer, M and Bogaert, G and Bondi, D and Bondu, F and Bonelli, L and Bonnand, R and Boschi, V and Bosi, L and Bouedo, T and Bradaschia, C and Branchesi, M and Briant, T and Brillet, A and Brisson, V and Bulik, T and Bulten, H J and Buskulic, D and Buy, C and Cagnoli, G and Calloni, E and Campeggi, C and Canuel, B and Carbognani, F and Cavalier, F and Cavalieri, R and Cella, G and Cesarini, E and Mottin, E Chassande- and Chincarini, A and Chiummo, A and Chua, S and Cleva, F and Coccia, E and Cohadon, P-F and Colla, A and Colombini, M and Conte, A and Coulon, J-P and Cuoco, E and Dalmaz, A and D’Antonio, S and Dattilo, V and Davier, M and Day, R and Debreczeni, G and Degallaix, J and Deléglise, S and Pozzo, W Del and Dereli, H and Rosa, R De and Fiore, L Di and Lieto, A Di and Virgilio, A Di and Doets, M and Dolique, V and Drago, M and Ducrot, M and Endrőczi, G and Fafone, V and Farinon, S and Ferrante, I and Ferrini, F and Fidecaro, F and Fiori, I and Flaminio, R and Fournier, J-D and Franco, S and Frasca, S and Frasconi, F and Gammaitoni, L and Garufi, F and Gaspard, M and Gatto, A and Gemme, G and Gendre, B and Genin, E and Gennai, A and Ghosh, S and Giacobone, L and Giazotto, A and Gouaty, R and Granata, M and Greco, G and Groot, P and Guidi, G M and Harms, J and Heidmann, A and Heitmann, H and Hello, P and Hemming, G and Hennes, E and Hofman, D and Jaranowski, P and Jonker, R J G and Kasprzack, M and Kéfélian, F and Kowalska, I and Kraan, M and Królak, A and Kutynia, A and Lazzaro, C and Leonardi, M and Leroy, N and Letendre, N and Li, T G F and Lieunard, B and Lorenzini, M and Loriette, V and Losurdo, G and Magazzù, C and Majorana, E and Maksimovic, I and Malvezzi, V and Man, N and Mangano, V and Mantovani, M and Marchesoni, F and Marion, F and Marque, J and Martelli, F and Martellini, L and Masserot, A and Meacher, D and Meidam, J and Mezzani, F and Michel, C and Milano, L and Minenkov, Y and Moggi, A and Mohan, M and Montani, M and Morgado, N and Mours, B and Mul, F and Nagy, M F and Nardecchia, I and Naticchioni, L and Nelemans, G and Neri, I and Neri, M and Nocera, F and Pacaud, E and Palomba, C and Paoletti, F and Paoli, A and Pasqualetti, A and Passaquieti, R and Passuello, D and Perciballi, M and Petit, S and Pichot, M and Piergiovanni, F and Pillant, G and Piluso, A and Pinard, L and Poggiani, R and Prijatelj, M and Prodi, G A and Punturo, M and Puppo, P and Rabeling, D S and Rácz, I and Rapagnani, P and Razzano, M and Re, V and Regimbau, T and Ricci, F and Robinet, F and Rocchi, A and Rolland, L and Romano, R and Rosińska, D and Ruggi, P and Saracco, E and Sassolas, B and Schimmel, F and Sentenac, D and Sequino, V and Shah, S and Siellez, K and Straniero, N and Swinkels, B and Tacca, M and Tonelli, M and Travasso, F and Turconi, M and Vajente, G and van Bakel, N and van Beuzekom, M and van den Brand, J F J and Van Den Broeck, C and van der Sluys, M V and van Heijningen, J and Vasúth, M and Vedovato, G and Veitch, J and Verkindt, D and Vetrano, F and Viceré, A and Vinet, J-Y and Visser, G and Vocca, H and Ward, R and Was, M and Wei, L-W and Yvert, M and żny, A Zadro and Zendri, J-P},
   year={2014},
   month=dec, pages={024001} }

@article{Yahya_2015,
   title={Cosmological performance of SKA HI galaxy surveys},
   volume={450},
   ISSN={1365-2966},
   url={http://dx.doi.org/10.1093/mnras/stv695},
   DOI={10.1093/mnras/stv695},
   number={3},
   journal={Monthly Notices of the Royal Astronomical Society},
   publisher={Oxford University Press (OUP)},
   author={Yahya, S. and Bull, P. and Santos, M. G. and Silva, M. and Maartens, R. and Okouma, P. and Bassett, B.},
   year={2015},
   month=may, pages={2251–2260} }

@article{Obreschkow_2009,
   title={A VIRTUAL SKY WITH EXTRAGALACTIC H I AND CO LINES FOR THE SQUARE KILOMETRE ARRAY AND THE ATACAMA LARGE MILLIMETER/SUBMILLIMETER ARRAY},
   volume={703},
   ISSN={1538-4357},
   url={http://dx.doi.org/10.1088/0004-637X/703/2/1890},
   DOI={10.1088/0004-637x/703/2/1890},
   number={2},
   journal={The Astrophysical Journal},
   publisher={American Astronomical Society},
   author={Obreschkow, D. and Klöckner, H.-R. and Heywood, I. and Levrier, F. and Rawlings, S.},
   year={2009},
   month=sep, pages={1890–1903} }

@article{Chen:2023wpj,
    author = "Chen, Anson and Gray, Rachel and Baker, Tessa",
    title = "{Testing the nature of gravitational wave propagation using dark sirens and galaxy catalogues}",
    eprint = "2309.03833",
    archivePrefix = "arXiv",
    primaryClass = "gr-qc",
    doi = "10.1088/1475-7516/2024/02/035",
    journal = "JCAP",
    volume = "02",
    pages = "035",
    year = "2024"
}

@misc{MeerKLASS,
      title={MeerKLASS: MeerKAT Large Area Synoptic Survey}, 
      author={Mario G. Santos and Michelle Cluver and Matt Hilton and Matt Jarvis and Gyula I. G. Jozsa and Lerothodi Leeuw and Oleg Smirnov and Russ Taylor and Filipe Abdalla and Jose Afonso and David Alonso and David Bacon and Bruce A. Bassett and Gianni Bernardi and Philip Bull and Stefano Camera and H. Cynthia Chiang and Sergio Colafrancesco and Pedro G. Ferreira and Jose Fonseca and Kurt van der Heyden and Ian Heywood and Kenda Knowles and Michelle Lochner and Yin-Zhe Ma and Roy Maartens and Sphesihle Makhathini and Kavilan Moodley and Alkistis Pourtsidou and Matthew Prescott and Jonathan Sievers and Kristine Spekkens and Mattia Vaccari and Amanda Weltman and Imogen Whittam and Amadeus Witzemann and Laura Wolz and Jonathan T. L. Zwart},
      year={2017},
      eprint={1709.06099},
      archivePrefix={arXiv},
      primaryClass={astro-ph.CO},
      url={https://arxiv.org/abs/1709.06099}, 
}

@article{Jolicoeur_2021,
   title={Detecting the relativistic bispectrum in 21cm intensity maps},
   volume={2021},
   ISSN={1475-7516},
   url={http://dx.doi.org/10.1088/1475-7516/2021/06/039},
   DOI={10.1088/1475-7516/2021/06/039},
   number={06},
   journal={Journal of Cosmology and Astroparticle Physics},
   publisher={IOP Publishing},
   author={Jolicoeur, Sheean and Maartens, Roy and De Weerd, Eline M. and Umeh, Obinna and Clarkson, Chris and Camera, Stefano},
   year={2021},
   month=jun, pages={039} }

@misc{Palmese:2025zku,
      title={Gravitational Wave Cosmology}, 
      author={Antonella Palmese and Simone Mastrogiovanni},
      year={2025},
      eprint={2502.00239},
      archivePrefix={arXiv},
      primaryClass={astro-ph.CO},
      url={https://arxiv.org/abs/2502.00239}, 
}

\appendix
\section{Amplitudes}\label{sec:amplitudes}
In this section, we show the number counts fluctuation for both redshift tracers and luminosity distance tracers. First we write out the general expression again for clarity:
\bea
\Delta(\boldsymbol{n},X) &=& b\delta_n+A^X_{GRV}\partial_r(\bm v\cdot\bm n) +\int_0^{\bar r} \d r\ \frac{A^X_{L}}{\bar r}\Delta_{\Omega}(\Phi+\Psi) \nn\\
&& +A_D^X(\bm v\cdot\bm n)+ g^{X}(\Psi,\Phi,r) \, . \
\eea
Then, we list the amplitudes for a redshift space tracer like galaxies or \hi\ intensity mapping, i.e. $X=z$:
\bea
A_{GRV} &=& A_{\rm{RSD}} = -\frac{1}{\mathcal{H}} \, , \\
A_D &=& \frac{5s-2}{\bar{r}\mathcal{H}} - 5s +b_e -\frac{\mathcal{H}'}{\mathcal{H}^2} \, , \\
A_L &=& \frac1{2} (5s-2)\frac{\bar{r}-r}{r} \, .
\eea
We note that $\bar{r}$ is the source's position and $r$ the integral's comoving distance, and $s$ and $b_e$ the magnification and evolution biases.

For a tracer living in luminosity distance space such as GWs, the amplitudes of the corrections to the number counts instead read:
\bea
A_{GRV} &=& A_{LSD} = -\ \frac{2\gamma}{\mathcal{H}} \label{eq:ALSD}\, ,\\
A_D &=& 1-2(\gamma+\beta) \label{eq:AD}\, ,\\
A_{L} &=& \frac12 \left[\left(\frac{\bar r-r}{r}\right)(\beta-2)+\frac{1}{1+\bar r \mathcal{H}}\right]\,\ \label{eq:AL} .
\eea
Here we define $\gamma \equiv \bar{r}\mathcal{H}/(1+\bar{r}\mathcal{H})$, and 
\be\label{eq:beta}
\beta\equiv
1-5s^{GW}+\gamma\left[\frac2{\bar r \mathcal{H}}+\gamma\left(\frac{\cal{H}'}{\mathcal{H}^2}-\frac1{\bar r\cal{H}}\right)-1-b_e^{GW}\right] \, ,
\ee
where $s^{GW}$ and $b_e^{GW}$ are the magnification and evolution biases, respectively.

\end{document}